\documentclass[1p,sort&compress]{elsarticle}
\pdfoutput=1   % required for arXiv submission if pdflatex is used
\usepackage{bm,amsfonts,amssymb,amsbsy}
\usepackage{graphicx}
\usepackage{color}
\usepackage{booktabs}
\usepackage{multirow}
\usepackage{multicol}
\usepackage{subfig}
\usepackage{placeins}
\newcommand{\nn}{\nonumber}

\newcommand{\Mvec}{\bm{M}}

\newcommand{\rvec}{\bm{r}}

\newcommand{\zvec}{\bm{z}}

\newcommand{\svec}{\bm{s}}
\newcommand{\sveca}{\bm{s}_a}

\newcommand{\wvec}{\bm{w}}

\newcommand{\Pvec}{\bm{P}}

\newcommand{\Rvec}{\bm{R}}
\newcommand{\dvec}{\bm{d}}

\newcommand{\xvec}{\bm{x}}
\newcommand{\nvec}{\bm{n}}

\newcommand{\ket}[1]{\vert #1\rangle}

\newcommand{\me}[3]{\langle #1\vert\ #2\ \vert #3\rangle}

\newcommand{\beq}{\begin{equation}}
\newcommand{\eeq}{\end{equation}}
\newcommand{\beqy}{\begin{eqnarray}}
\newcommand{\eeqy}{\end{eqnarray}}
\newcommand{\beqqy}{\begin{eqnarray*}}
\newcommand{\eeqqy}{\end{eqnarray*}}

\newcommand{\Ecm}{E_{\rm cm}}
\newcommand{\qcm}{\bm{q}_{\rm cm}}
\newcommand{\qcma}{\bm{q}_{{\rm cm},a}}
\newcommand{\Maple}{\textsc{Maple}}

\newcommand{\Ecmk}{E_{{\rm cm},k}}
\begin{document}
\begin{flushright}
MITP/17-044\\CP3-Origins-2017-27
\end{flushright}
\begin{frontmatter}
\title{
    Estimating the two-particle $K$-matrix 
    for multiple partial waves and decay channels 
    from finite-volume energies }

\author[cmu]{Colin~Morningstar}
\author[usden]{John~Bulava}
\author[cmu]{Bijit~Singha}
\author[cmu]{Ruair\'i~Brett}
\author[cmu]{Jacob~Fallica}
\author[pitt]{Andrew~Hanlon}
\author[mainz]{Ben~H\"{o}rz}

\address[cmu]{Department~of~Physics, Carnegie~Mellon~University, 
              Pittsburgh, PA~15213, USA}
\address[usden]{Dept.~of Mathematics~and~Computer~Science and CP3-Origins, 
                University of Southern Denmark, Campusvej 55, 5230 Odense M, Denmark}
\address[pitt]{Dept.~of Physics and Astronomy, University of Pittsburgh, 
               Pittsburgh, PA 15260, USA}
\address[mainz]{PRISMA Cluster of Excellence and Institute for Nuclear Physics, 
                University of Mainz, Johann Joachim Becher-Weg 45, 55099 Mainz, Germany}

\begin{abstract}
An implementation of estimating the two-to-two $K$-matrix from finite-volume
energies based on the L\"{u}scher formalism and involving a Hermitian matrix
known as the ``box matrix'' is described.  The method includes higher 
partial waves and multiple decay channels.  Two fitting procedures for
estimating the $K$-matrix parameters, which properly incorporate all statistical 
covariances, are discussed.   Formulas and software for handling total spins up 
to $S=2$ and orbital angular momenta up to $L=6$ are obtained for total momenta 
in several directions.  First tests involving $\rho$-meson decay to two pions
include the $L=3$ and $L=5$ partial waves, and the contributions from these
higher waves are found to be negligible in the elastic energy range.
\end{abstract}

\begin{keyword}
lattice QCD, meson scattering, baryon scattering
	\PACS{% 
12.38.Gc\sep %Lattice QCD calculations
11.15.Ha\sep %Lattice gauge theory
11.30.Rd\sep %Chiral symmetries
12.38.Aw\sep %General properties of QCD
13.30.Eg\sep %Hadronic decays
13.75.Lb\sep %meson-meson interactions 
13.85.Dz\sep %elastic scattering
14.40.Be\sep %Light mesons (S=C=B=0)
14.65.Bt %Light quarks
}    %masses and mixing (electroweak interactions)
\end{keyword}
\end{frontmatter}

\section{Introduction}
Currently, the best method of computing the masses and other properties 
of hadrons from quantum chromodynamics (QCD) involves estimating the QCD 
path integrals using Markov-chain Monte Carlo methods, which requires
formulating the theory on a space-time lattice.  Such calculations
are necessarily carried out in finite volume.  However, most of the 
excited hadrons we seek to study are unstable resonances.  Fortunately,
it is possible to deduce the masses and widths of resonances from
the spectrum determined in finite volume.

The idea that finite-volume energies can be related to 
infinite-volume scattering processes is actually rather old, dating back to 
Refs.~\cite{Reifman:1955ca,DeWitt:1956be} in the mid-1950s.  First
suggestions of applying such techniques for gauge field theories appeared
in Ref.~\cite{Luscher:1983rk}.  In Ref.~\cite{Luscher:1985dn} in 1986, L\"uscher
studied the volume dependence of the energy spectrum of stable particle
states in massive quantum field theories, then examined the volume
dependence of scattering states in Ref.~\cite{Luscher:1986pf} soon thereafter.
In Ref.~\cite{Luscher:1990ux} in 1991, L\"uscher then found relationships
between finite-volume energies and infinite-volume scattering phase
shifts in the case of two identical spinless particles having zero total 
momentum and interacting via a central potential.  The advantages of using 
sectors with non-zero total momenta were then described by Rummukainen and 
Gottlieb in Ref.~\cite{Rummukainen:1995vs} in 1995.  These calculations
were later revisited in 2005 using a field theoretic approach
by Kim, Sachrajda, and Sharpe in Ref.~\cite{Kim:2005gf}.  This work focused 
on the case of a single channel of identical spinless particles, but the
total momentum could be any value allowed by the boundary conditions.
The results of Ref.~\cite{Kim:2005gf} were eventually generalized in 
Refs.~\cite{Luu:2011ep,Fu:2011xz,Leskovec:2012gb,Hansen:2012tf,Gockeler:2012yj,Briceno:2014oea}, 
among others, to treat multi-channels with different particle masses and 
nonzero spins.

As new techniques, such as those introduced in Refs.~\cite{Peardon:2009gh} and
\cite{Morningstar:2011ka}, have made possible accurate determinations of the 
energies of states created by two-hadron operators, an increasing number
of lattice QCD studies of scattering phase shifts have 
appeared\cite{Bulava:2016mks,Alexandrou:2017mpi,Bali:2015gji,Fu:2016itp,Feng:2014gba,
Feng:2010es,Detmold:2015qwf,Berkowitz:2015eaa,Orginos:2015aya,Beane:2011sc,
Pelissier:2012pi,Aoki:2011yj,Moir:2016srx,Briceno:2016mjc,Briceno:2016kkp,
Dudek:2016cru,Wilson:2015dqa,Wilson:2014cna,Dudek:2012xn,Dudek:2012gj,Lang:2016jpk,
Lang:2015hza,Lang:2014yfa,Mohler:2013rwa,Prelovsek:2013ela,Mohler:2012na,Lang:2012sv,
Lang:2011mn,Guo:2016zos,Helmes:2017smr,Liu:2016cba,Helmes:2015gla}.

The purpose of this paper is to provide explicit formulas, software, and new fitting
implementations for carrying out two-particle scattering studies using energies
obtained from lattice QCD.
We first express the quantization condition that relates lattice QCD energies
to the scattering matrix in terms of the more convenient $K$-matrix and a
Hermitian matrix $B$ which we refer to as the ``box matrix'' since it essentially
describes how the spherical partial waves manage to fit into a cubic finite volume.
We restrict our attention to periodic boundary conditions.  
We obtain explicit results for several spins and center-of-momentum orbital angular 
momenta up to $L=6$ with total momentum of the form 
$\Pvec=(0,0,0), (0,0,p), (0,p,p), (p,p,p)$ for $p>0$, many of which have
never before appeared in the literature.  More importantly,
the software for evaluating all of these box matrix elements is made available.
For multiple partial waves and channels, the quantization condition at an allowed
energy is insufficient to determine the entire $K$-matrix at that energy.
Approximating the $K$-matrix elements using physically-motivated functions of
the energy involving a handful of parameters is needed, with the hopes that
these parameters can be estimated by appropriate fits using a sufficiently
large number of different energies. We discuss two fitting strategies
with such a goal in mind.  In our first tests of the fitting methods, we 
incorporate the $L=3$ and $L=5$ partial waves in the decay of the $\rho$-meson 
to two pions and find their contributions to be negligible in the elastic 
energy region.

\section{Quantization condition}

We begin by summarizing the quantization condition that relates each
finite-volume energy to the scattering $S$-matrix.
We work in an $L^3$ spatial volume with periodic boundary conditions.
For a total momentum $\Pvec=(2\pi/L)\dvec$,
where $\dvec$ is a vector of integers, we determine the total energy $E$
in the lab frame for a particular two-particle interacting state
in our lattice QCD simulations.   We boost to the 
center-of-momentum frame and calculate
\beq
   \Ecm = \sqrt{E^2-\Pvec^2},\qquad \gamma =\frac{E}{\Ecm}.
\eeq
Let $N_d$ denote the number of two-particle channels that are open,
and denote the masses and spins of the two scattering particles in
channel $a$ by $m_{ja}$ and $s_{ja}$, respectively, for $j=1,2$.
If $\vert\Ecm^2\vert \geq \vert m_{1a}^2-m_{2a}^2\vert$, then 
we can calculate the following quantities in each channel:
\begin{eqnarray}
   \qcma^{2} &=& \frac{1}{4} \Ecm^{2}
   - \frac{1}{2}(m_{1a}^2+m_{2a}^2) + \frac{(m_{1a}^2-m_{2a}^2)^2}{4\Ecm^{2}},\\
   u_a^2&=& \frac{L^2\qcma^2}{(2\pi)^2},\qquad
 \sveca= \left(1+\frac{(m_{1a}^2-m_{2a}^2)}{\Ecm^{2}}\right)\dvec.
\end{eqnarray}
The total energy $E$ is then related to the dimensionless unitary scattering $S$-matrix
through the quantization condition\cite{Luscher:1990ux,Rummukainen:1995vs,Kim:2005gf,Briceno:2014oea}:
\beq
   \det[1+F^{(\Pvec)}(S-1)]=0.
\label{eq:quant1}
\eeq
We use an orthonormal basis of states, each labelled by $\ket{Jm_JLS a}$, where $J$ is
the total angular momentum of the two particles in the center-of-momentum frame, $m_J$ is 
the projection of the total angular momentum onto the $z$-axis, $L$ is the orbital angular 
momentum of the two particles in the center-of-momentum frame (not to be confused with the
lattice length here), $S$ in the basis vector is 
the total spin of the two particles (not the scattering matrix).  The angular 
momentum, orbital angular momentum, and spin operators are defined as usual in terms of the
Pauli-Lubanski tensor $ W^\mu = -\textstyle\frac{1}{2}\epsilon^{\mu\nu\alpha\beta}
M_{\nu\alpha}P_\beta,$ where $P^\mu$ generates space-time shifts and
the antisymmetric tensor $M$ combines generators of rotations and Lorentz boosts.
The index $a$ is generalized to refer to species, the spins $s_1,s_2$, intrinsic 
parities $\eta^P_1,\eta^P_2$, isospins $I_1, I_2$, isospin projections $I_{z1}, I_{z2}$, 
and possibly $G$-parities $\eta_1^G, \eta_2^G$ of particle 1 and 2. 
Other quantum numbers, such as strangeness and charm, could also be
included.  The $F^{(\Pvec)}$ matrix in this basis is given by
\beqy
&&\langle  J'm_{J'}L'S'a'\vert F^{(\Pvec)}\vert Jm_JLSa\rangle
=\delta_{a'a}\delta_{S'S}\ \frac{1}{2}
\Bigl\{\delta_{J'J}\delta_{m_{J'}m_J}\delta_{L'L}\nn\\
&&\qquad + \langle J'm_{J'}\vert L'm_{L'} Sm_{S}\rangle
\langle Lm_L Sm_S\vert Jm_J\rangle
W_{L'm_{L'};\ Lm_L}^{(\Pvec a)}\Bigr\},
\eeqy
where $\langle j_1m_1 j_2m_2\vert JM\rangle$ are the familiar Clebsch-Gordan coefficients,
and the $W^{(\Pvec a)}$ matrix elements are given by
\beqy
-iW^{(\Pvec a)}_{L'm_{L'};\ Lm_L} 
&=& \sum_{l=\vert L'-L\vert}^{L'+L}\sum_{m=-l}^l
   \frac{ {\cal Z}_{lm}(\svec_a,\gamma,u_a^2) }{\pi^{3/2}\gamma u_a^{l+1}}
\sqrt\frac{(2{L'}+1)(2l+1)}{(2L+1)}\nonumber\\
&&\qquad\qquad\times \langle {L'} 0,l 0\vert L 0\rangle
\langle {L'} m_{L'},  l m\vert  L m_L\rangle.
\label{eq:Wdef2}
\eeqy
The Rummukainen-Gottlieb-L\"uscher (RGL) shifted zeta functions 
${\cal Z}_{lm}$, introduced in Refs.~\cite{Luscher:1990ux,Rummukainen:1995vs},
are evaluated using \cite{Luu:2011ep,Gockeler:2012yj}
\beqy
   {\cal Z}_{lm}(\svec,\gamma,u^2)&=&\sum_{\nvec\in \mathbb{Z}^3}
  \frac{{\cal Y}_{lm}(\zvec)}{(\zvec^2-u^2)}e^{-\Lambda(\zvec^2-u^2)}
 +\delta_{l0}\frac{\gamma\pi}{\sqrt{\Lambda}} F_0(\Lambda u^2)
\nn\\
 &+&\frac{i^l\gamma}{\Lambda^{l+1/2}} \int_0^1\!\!dt 
\left(\frac{\pi}{t}\right)^{l+3/2}\! e^{\Lambda t u^2}
\sum_{\nvec\in \mathbb{Z}^3\atop \nvec\neq 0}
e^{\pi i \nvec\cdot\svec}{\cal Y}_{lm}(\wvec)
\  e^{-\pi^2\wvec^2/(t\Lambda)},
\label{eq:zaccfinal2}
\eeqy
with ${\cal Y}_{lm}(\xvec)=\vert \xvec\vert^l\ Y_{lm}(\widehat{\xvec})$ and
\beqy
 && \zvec= \nvec -\gamma^{-1} \bigl[\textstyle\frac{1}{2}
+(\gamma-1)s^{-2}\nvec\cdot\svec \bigl]\svec,\\
&& \wvec=\nvec - (1  - \gamma) s^{-2}
 \svec\cdot\nvec\svec,\\
  &&   F_0(x) =  -1+\frac{1}{2}
\int_0^1\!\! dt\ \frac{e^{tx}-1 }{t^{ 3/2}}.
\eeqy
We choose $\Lambda\approx 1$ for fast convergence of the summation,
and the integral in Eq.~(\ref{eq:zaccfinal2}) is efficiently done
using Gauss-Legendre quadrature.  The function $F_0(x)$ is given in 
terms of the Dawson or erf function:
\beq
 F_0(x)=  \left\{\begin{array}{ll}
  e^{x}\left( 2\sqrt{x} D(\sqrt{x})-1 \right), &  (x\geq 0),\\
  -e^{x}-\sqrt{-x\pi}\ {\rm erf}(\sqrt{-x}), & (x<0).
 \end{array}\right.
\eeq
Many of these RGL shifted zeta functions are related to one another.
Some of these relations and the symmetry properties of these functions
are summarized in Appendix~A.

The above relations apply for both distinguishable and indistinguishable
particles, since the associated symmetry factors cancel in the quantization
condition in general\cite{Hansen:2012tf}.  The only difference that occurs
with indistinguishable particles is that certain combinations of $L$ and $S$
cannot occur.  In the absence of isospin, $L+S$ must be even for identical
particles.  For identical particles of isospin $I_1$,  $L+S+I-2I_1$ must be 
even, where $I$ is the total isospin.

\section{The $K$-matrix and box matrix}

The quantization condition in Eq.~(\ref{eq:quant1}) is a single relation
between an energy $E$ determined in finite-volume and the entire
$S$-matrix.  In the very limited case of a single channel with a 
single partial wave, this relationship can be used to directly extract the 
scattering phase shift at energy $E$.  When multiple partial waves or multiple
channels are involved, this single relation is clearly not sufficient
to extract all of the $S$-matrix elements at energy $E$.  The best way to 
proceed is to approximate the $S$-matrix elements using physically motivated 
functions of the energy $E$ involving a handful of parameters.  Values
of these parameters can then hopefully be estimated by appropriate fits
using a sufficiently large number of different energies.  Such a procedure
has long been standard in the partial wave analyses of particle experiments.

The $S$-matrix in Eq.~(\ref{eq:quant1}) is dimensionless and unitary.
Since it is easier to parametrize a real symmetric matrix than a unitary 
matrix, one usually employs the real and symmetric $K$-matrix.  
The $K$-matrix was first introduced by Wigner\cite{Wigner:1946zz} and 
Wigner and Eisenbud\cite{Wigner:1947zz} studying resonances in nuclear
reactions. Its first use in particle physics was an analysis of 
resonance production in $Kp$ scattering by Dalitz and Tuan\cite{Dalitz:1960du}.
A comprehensive review of the $K$-matrix formalism is given in 
Ref.~\cite{Badalian:1981xj}, and a more recent review appears in 
Ref.~\cite{Chung:1995dx}.  Recent applications of the $K$-matrix to study 
resonances in lattice QCD can be found in 
Refs.~\cite{Wilson:2014cna,Dudek:2016cru}.

Defining the transition operator $T$ via $S=1+iT$, the $K$-matrix can be 
defined by
\beq
  K = (2T^{-1}+i)^{-1},\qquad K^{-1}=2T^{-1}+i,
\eeq
and hence, its relationship to the $S$-matrix is 
\beq
   S = (1+iK)(1-iK)^{-1} = (1-iK)^{-1}(1+iK). 
\eeq
Using the above definition, it is straightforward to show that the
unitarity of $S$ implies the Hermiticity of $K$.
Time reversal invariance of the $S$-matrix implies that the $K$-matrix 
must be symmetric.  Given its Hermiticity, this means that $K$ is 
real and symmetric.  Rotational invariance implies that the $K$-matrix 
must have the form
\beq 
  \langle J'm_{J'}L^\prime S^\prime a'\vert\ K
\ \vert Jm_JLS  a\rangle = \delta_{J'J}\delta_{m_{J'}m_J}
 \ K^{(J)}_{L'S'a';\ LS a}(\Ecm),
\label{eq:Kbasis}
\eeq
where $a',a$ denote other defining quantum numbers, such as channel, and
$K^{(J)}$ is a real, symmetric matrix that is independent of 
$m_J$. Invariance under parity also gives us that
\beq
   K^{(J)}_{L'S'a';\ LSa}(\Ecm)=0\quad\mbox{when 
  $\eta^{P\prime}_{1a'}\eta^P_{1a}\eta^{P\prime}_{2a'}\eta^P_{2a}(-1)^{L'+L}=-1$},
\label{eq:Kparity}
\eeq
where $\eta_{ja}^P$ denotes the intrinsic parity of particle $j$ in
the channel associated with $a$.

For a single channel of spinless particles, the $S$-matrix is diagonal
with elements $S_L=e^{2i\delta_L}$, so the
$K$-matrix diagonal elements are $ K_L = \tan\delta_L,$ where $\delta_L$ is the 
scattering phase shift of the $L$-th partial wave.
For a short-ranged potential in the single-channel case, one can derive the 
effective range expansion, which states that
\beq
  q_{\rm cm}^{2L+1}\cot\delta_L(q_{\rm cm})=q_{\rm cm}^{2L+1}K^{-1}_L
=\sum_{n=0}c_{2n}q_{\rm cm}^{2n}
=-\frac{1}{a_L}+\frac{r_L}{2}q_{\rm cm}^2+O(q_{\rm cm}^4),
\eeq
where $q_{\rm cm}=\sqrt{\qcm^2}$,
the constants $a_L$ are called scattering lengths (although
only the $S$-wave constant $a_0$ has the dimensions of a length), and $r_L$
are known as the effective ranges.
The multichannel generalization\cite{Ross:1961aa,deSwart:1962aa,Burke:2011}
of the effective range expansion is
\beq
 K^{-1}_{L'S'a';\ LSa}(\Ecm)=q_{{\rm cm},a'}^{-L'-\frac{1}{2}}
 \ {\widehat{K}}^{-1}_{L'S'a';\ LSa}(\Ecm)
  \ q_{{\rm cm},a}^{-L-\frac{1}{2}},
\label{eq:Keffrange}
\eeq
where ${\widehat{K}}^{-1}_{L'S'a';\ LSa}(\Ecm)$ is a real, symmetric, and analytic 
function of the center-of-momentum energy $\Ecm$.

The effective range expansion given in Eq.~(\ref{eq:Keffrange}) suggests the convenience
of writing
\beq
 K^{-1}_{L'S'a';\ LSa}(\Ecm)=u_{a'}^{-L'-\frac{1}{2}}\ {\widetilde{K}}^{-1}_{L'S'a';\ LSa}(\Ecm)
  \ u_a^{-L-\frac{1}{2}},
\eeq
since ${\widetilde{K}}^{-1}_{L'S'a';\ LSa}(\Ecm)$ is real and symmetric and expected to behave
smoothly with energy $\Ecm$.  It is then straightforward to show that the quantization
condition of Eq.~(\ref{eq:quant1}) can be written
\beq
\det(1-B^{(\Pvec)}\widetilde{K})=\det(1-\widetilde{K}B^{(\Pvec)})=0,
\label{eq:quant2}
\eeq
where we define the \textit{box matrix} by
\beqy
 && \me{J'm_{J'}L'S'a'}{B^{(\Pvec)}}{Jm_JLS a} =
-i\delta_{a'a}\delta_{S'S} \ u_a^{L'+L+1}\ W_{L'm_{L'};\ Lm_L}^{(\Pvec a)}  \nn\\
&&\qquad\qquad \times\langle J'm_{J'}\vert L'm_{L'},Sm_{S}\rangle
\langle Lm_L,Sm_S\vert Jm_J\rangle.
\label{eq:Bmatdef}
\eeqy
This box matrix $B^{(\Pvec)}$ is Hermitian for $u_a^2$ real.  Whenever
$\det \widetilde{K}\neq 0$, which is usually true in the presence of interactions,
the quantization condition can also be written
\beq
  \det(\widetilde{K}^{-1}-B^{(\Pvec)})=0.
\label{eq:quant3}
\eeq
The Hermiticity of $B^{(\Pvec)}$ and the fact that $\widetilde{K}$ is
real and symmetric for real $u_a^2$ ensures that the determinants in the 
quantization conditions of Eqs.~(\ref{eq:quant2}) and (\ref{eq:quant3}) 
are real.  Note that $\widetilde{K}$ and $B^{(\Pvec)}$ do not commute in
general, which means $1-B^{(\Pvec)}\widetilde{K}$ and 
$1-\widetilde{K}B^{(\Pvec)}$ are not Hermitian.  However, it is easy
to show that their determinants must be real.

Again, rotational invariance of the $K$-matrix implies that $\widetilde{K}$ has the form
\beq 
  \langle J'm_{J'}L^\prime S^\prime a'\vert\ \widetilde{K}
\ \vert Jm_JLS  a\rangle = \delta_{J'J}\delta_{m_{J'}m_J}
 \ {\cal K}^{(J)}_{L'S'a';\ LS a}(\Ecm).
\label{eq:Ksmooth}
\eeq
When $S=S'=0$, then $J=L$ and $J'=L'$ yielding
\beq 
  \langle J'm_{J'}L^\prime 0 a'\vert\ \widetilde{K}
\ \vert Jm_JL0  a\rangle = \delta_{J'J}\delta_{m_{J'}m_J}\delta_{J'L'}\delta_{JL}
 \ {\cal K}^{(L)}_{a';\ a}(\Ecm).
\label{eq:Ksmoothb}
\eeq
Given that $\widetilde{K}^{-1}$ is expected to be analytic in $\Ecm$,
an obvious parametrization of the inverse of the $\widetilde{K}$-matrix 
over a small range of energies is using a symmetric matrix of polynomials 
in $\Ecm$:
\beq
 {\cal K}^{(J)-1}_{\alpha\beta}(\Ecm) = \sum_{k=0}^{N_{\alpha\beta}} 
c_{\alpha\beta}^{(Jk)} \Ecm^k,
\label{eq:Kfit1}
\eeq
where $\alpha,\beta$ are compound indices referring to orbital momentum $L$, total spin $S$, and
channel $a$, and the $c_{\alpha\beta}^{(Jk)}$ form a real symmetric matrix for each $k$.
Another common parame\-trization (see, for example, Ref.~\cite{Chung:1995dx})
expresses the $\widetilde{K}$-matrix as a sum of poles with a 
background described by a symmetric matrix of polynomials:
\beq
  {\cal K}^{(J)}_{\alpha\beta}(\Ecm) = \sum_{p} \frac{g_\alpha^{(Jp)} g_\beta^{(Jp)}}{\Ecm^2-m_{Jp}^2}
     + \sum_k d_{\alpha\beta}^{(Jk)} \Ecm^k,
\label{eq:Kfit2}
\eeq
where the couplings $g_\alpha^{(Jp)}$ are real and the background coefficients
$d_{\alpha\beta}^{(Jk)}$ form a real symmetric matrix for each $k$.  These can be written
in Lorentz invariant form using $\Ecm=\sqrt{s}$, where the Mandelstam variable $s=(p_1+p_2)^2$,
with $p_j$ being the four-momentum of particle $j$.

\section{Block diagonalization}

In the previous sections, we expressed the matrices $F^{(\Pvec)}$ and $B^{(\Pvec)}$
in terms of the orthonormal center-of-momentum frame basis states labelled by $\ket{Jm_JLS a}$.  
In this basis, the quantization condition in each of Eqs.~(\ref{eq:quant1}), (\ref{eq:quant2}) and
(\ref{eq:quant3}) is problematic due to the need to evaluate the determinant of an 
infinite matrix. If we can transform to a basis in which both $B^{(\Pvec)}$ and $\widetilde{K}$ 
are block diagonal, then we only need to examine the determinant separately in each block.
Each block has infinite dimension, but if we truncate in the orbital angular momentum, 
keeping only states with $L\leq L_{\rm max}$, then each truncated block has a finite and 
reasonably small size.

Under a symmetry transformation $G$ which is either an ordinary spatial rotation $R$ or 
spatial inversion $I_s$, the total momentum $\Pvec$ changes to $G\Pvec$, and if we define a 
unitary matrix $Q^{(G)}$ by
\beq
      \me{J'm_{J'}L'S'a'}{Q^{(G)}}{ Jm_JLS a}
   = \left\{\begin{array}{ll}
 \delta_{J'J}\delta_{L'L}\delta_{S'S}\delta_{a'a}
     D^{(J)}_{m_{J'}m_{J}}(R),  & (G=R),\\[4pt]
  \delta_{J'J}\delta_{m_{J'}m_J}\delta_{L'L}\delta_{S'S}\delta_{a'a}
     (-1)^{L}, & (G=I_s),
 \end{array}\right.
\label{eq:Qsymdef}
\eeq
where $D^{(J)}_{m'm}(R)$ are the familiar Wigner rotation matrices,
one can show that the box matrix satisfies
\beq
   B^{(G\Pvec)} = Q^{(G)}\ B^{(\Pvec)}\ Q^{(G)\dagger}.
\label{eq:Brotate}
\eeq
The result in Eq.~(\ref{eq:Brotate}) is very important since it allows
us to block diagonalize the $B^{(\Pvec)}$ matrix.  If $G$ is an element of the 
little group of $\Pvec$, then $G\Pvec=\Pvec$ and $G\svec_a=\svec_a$,  and we 
have
\beq
   B^{(\Pvec)} = Q^{(G)}\ B^{(\Pvec)}\ Q^{(G)\dagger},
 \qquad\mbox{($G$ in little group of $\Pvec$).}
\label{eq:Breduce2}
\eeq
Since $Q^{(G)}$ is unitary, this implies that the $B^{(\Pvec)}$ matrix commutes 
with the matrix $Q^{(G)}$ for all $G$ in the little group of $\Pvec$.
This means that we can simultaneously diagonalize $B^{(\Pvec)}$ and $Q^{(G)}$.  
By rotating into a basis formed by the eigenvectors of $Q^{(G)}$, we can reduce 
the $B^{(\Pvec)}$ matrix into a block diagonal form since the matrix elements
of $B^{(\Pvec)}$ between different eigenvectors of $Q^{(G)}$ must vanish.

Rotations, reflections, and spatial inversion do not change $J,L,S,a$ when
acting on basis state $\ket{Jm_JLSa}$.  These symmetry operations only mix
states of different $m_J$. A partial diagonalization of $B^{(\Pvec)}$
can be achieved by diagonalizing $Q^{(G)}$ for each $J,L,S,a$, or
equivalently, by projecting onto the irreducible representations of 
the little group of $\Pvec$.  These eigenvectors or projections
can be labelled by the irreducible representation (irrep) $\Lambda$ and irrep 
row $\lambda$ of the little group, and an integer $n$ identifying each occurrence 
of the irrep $\Lambda$ in the $\ket{Jm_JLS a}$ reducible representation.
In other words, to block diagonalize $B^{(\Pvec)}$, we need to apply a particular
unitary change of basis:
\beq
  \ket{\Lambda\lambda n JLS a}= \sum_{m_J} c^{J\eta;\,\Lambda\lambda n}_{m_J} 
 \ket{Jm_JLS a},
 \label{eq:bdtrans}
\eeq
where $\eta=(-1)^L$.  From Eq.~(\ref{eq:Qsymdef}), one sees that the transformation 
coefficients depend on $J$ and $(-1)^L$, but not on $L$ itself, and are 
independent of $S,a$.  

Our procedure for computing the transformation coefficients is as follows.
To simplify notation, we suppress the $S,a$ indices, define $\eta = (-1)^L$,
and abbreviate the state $\ket{Jm_JLS a}$ by $\ket{J^\eta m_J}$.
For a given $J,L,S,a$, we apply the 
standard group theoretical projections onto the $(2J+1)$ basis vectors 
$\ket{J^\eta,J}, \ket{J^\eta,J-1}, \dots, \ket{J^\eta,-J}$
for the first row $\lambda=1$ of each irrep of the little group,
producing the vectors (one for each $m_J$)
\beq
     \frac{d_\Lambda}{g_{\cal G}}\sum_{m'}
   \sum_{G\in {\cal G}} \Gamma^{(\Lambda)\ast}_{\lambda\lambda}(G)
     \ Q^{(J\eta)}_{m'm_J}(G)\ \vert J^\eta m'\rangle,
\label{eq:Bproject}
\eeq
where ${\cal G}$ denotes the little group, $g_{\cal G}$ is the number
of elements in the little group, $d_\Lambda$ is the
dimension of the irrep $\Lambda$, $\Gamma^{(\Lambda)}(G)$
is the unitary matrix representing $G$ in the $\Lambda$ irrep,
and for rotations $R$ and spatial inversion $I_s$, we have
\beq
    Q^{(J\eta)}_{m'm}(R)=D^{(J)}_{m'm}(R),\qquad
    Q^{(J\eta)}_{m'm}(I_s)=\delta_{m'm}\eta.
\eeq
If the irrep $\Lambda$ does not occur in the $J$ representation of $SU(2)$ subduced to 
${\cal G}$, all of the resulting vectors will be zero.  If the irrep $\Lambda$ 
occurs once in the subduction, then only one nonzero vector will occur, 
which can then be suitably normalized.  If the $\Lambda$ irrep occurs more than
once, then there is some freedom in choosing the basis vectors.  We first look
for linear combinations of the projected vectors that are simpler in form
with less terms, order the resulting vectors by increasing complexity, 
then apply a Gram-Schmidt procedure to obtain orthonormal basis states.  Once 
we have basis vectors $\vert \Lambda\lambda n\rangle$ for the first row 
$\lambda=1$ of all $\Lambda$ irreps, where positive integer $n$ is the 
occurrence label, we then obtain the partner basis vectors for the other 
rows $\mu$ using the transfer operation
\beq
     \vert\Lambda\mu n\rangle = \frac{d_\Lambda}{g_{\cal G}}
   \sum_{G\in {\cal G}} \Gamma^{(\Lambda)\ast}_{\mu\lambda}(G)
   \ Q^{(G)}\vert \Lambda\lambda n\rangle.
\eeq

Our choices of irreducible representation matrices $\Gamma^{(\Lambda)}_{\mu\nu}(G)$ are presented
in Ref.~\cite{Morningstar:2013bda}, and the irrep labels for the various little groups
are listed in Ref.~\cite{Morningstar:2013bda} as well.  Given these choices, we have applied
the above procedure using software written in \Maple.  We have computed basis 
vectors for all $J\leq 8$ for momenta $\Pvec=(0,0,0), (0,0,p), (0,p,p), (p,p,p)$,
with $p>0$.  Given the very large number of such basis vectors, it is not possible
to present them all here.  However, explicit tables containing our basis vectors have been 
generated and are available from Ref.~\cite{gitavail}.

Expressing the box matrix in this basis, one can show that 
$B^{(\Pvec)}$ is diagonal in $\Lambda,\lambda$, but not in the occurrence 
index $n$. Given Eq.~(\ref{eq:Bmatdef}), we find that we can write
\beq
 \me{\Lambda'\lambda' n'J'L'S' a'}{B^{(\Pvec)}}{\Lambda\lambda nJLS a}
 = \delta_{\Lambda'\Lambda}\delta_{\lambda'\lambda}\delta_{S'S}
 \delta_{a'a}\ B^{(\Pvec\Lambda_B Sa)}_{J'L'n';\ JLn}(E).
\label{eq:Bbdform}
\eeq
The box matrix depends on $a$ only through $u_a$ and $\svec_a$.
Notice that in Eq.~(\ref{eq:Bbdform}) we use the irrep label $\Lambda_B$ instead of 
$\Lambda$ to label the matrix elements of $B^{(\Pvec)}$.  We wish to reserve the 
irrep $\Lambda$ to describe the symmetry of the block in question for the full system, 
which includes the intrinsic parities of the constituent particles.  The $B^{(\Pvec)}$ 
matrices transform independently of these intrinsic parities.  If $\eta^P_{1a}\eta^P_{2a}=1$,
then $\Lambda_B=\Lambda$, but for a channel in which $\eta^P_{1a}\eta^P_{2a}=-1$,
$\Lambda$ and $\Lambda_B$ may not be the same.  When the total momentum $\Pvec=0$, 
$\Lambda$ and $\Lambda_B$ have opposite parity labels: $g\leftrightarrow u$ in
the $O_h$ label subscripts.  For $\Pvec\neq 0$ when $\eta^P_{1a}\eta^P_{2a}=-1$,
one must carefully deduce the relationship between $\Lambda$ and $\Lambda_B$.
In each case, one consults the conjugacy classes of the little group and
identifies the classes that involve improper transformations (those involving
parity); for a given irrep $\Lambda$, one takes the character vector and
flips the signs of the characters in the classes involving improper symmetry
operations, and the resulting character vector identifies $\Lambda_B$.
The relationships of $\Lambda_B$ to $\Lambda$ when $\eta^P_{1a}\eta^P_{2a}=-1$
for various momenta $\Pvec$ are summarized in Table~\ref{tab:lambdaB}.  

\begin{table}
\caption[tab:lambdaB]{Relationship of $B$-matrix irrep $\Lambda_B$ to full symmetry
  irrep $\Lambda$.  When the product of intrinsic parities $\eta^P_{1a}\eta^P_{2a}=1$, then
 $\Lambda_B=\Lambda$.  For a channel in which  $\eta^P_{1a}\eta^P_{2a}=-1$, the
relationship of $\Lambda_B$ to $\Lambda$ is shown in the table below.  LG denotes
``little group'', and $n>0$ in $\Pvec=(2\pi/L)\dvec$ below.  For details about the 
little group irreps (single-valued and double-valued) listed below, 
see Ref.~\cite{Morningstar:2013bda}.
\label{tab:lambdaB}}
\begin{center}
\begin{tabular}{|c|c|l|} \hline
$\dvec$ & LG & $\Lambda_B$ relationship to $\Lambda$ when $\eta^P_{1a}\eta^P_{2a}=-1$\\ \hline\hline
$(0,0,0)$ & $O_h$    &   Subscript $g\leftrightarrow u$ \\
$(0,0,n)$ & $C_{4v}$ &   $A_1 \leftrightarrow A_2;\  B_1 \leftrightarrow B_2$;  $E, G_1, G_2$ unchanged\\
$(0,n,n)$ & $C_{2v}$ &   $A_1 \leftrightarrow A_2;\  B_1 \leftrightarrow B_2$;  $G$ unchanged\\
$(n,n,n)$ & $C_{3v}$ &   $A_1 \leftrightarrow A_2;\  F_1 \leftrightarrow F_2$;  $E, G$ unchanged\\ \hline
\end{tabular}
\end{center}
\end{table}

With these transformation coefficients, the box matrix elements can be obtained in 
this block diagonal basis by straightforward matrix multiplication.
\Maple\ code was written to perform these evaluations.  In this software,
the symmetries of the RGL zeta functions are used to express the results 
in terms of the minimum number of independent functions (those listed in 
Tables~\ref{tab:zsymrest} to \ref{tab:zsymcd}). To check our results,
we explicitly verified that (a) elements of the box matrix between
states in different blocks are all zero, (b) the box matrix in each
block is Hermitian, and (c) the box matrix blocks are exactly the same 
for all rows in each irrep. 

Expressions in terms of the RGL shifted zeta functions for a small 
selection of the box matrix elements 
$B^{(\Pvec\Lambda_B S a)}_{J'L'n';\ JLn}(E)$ which we have determined
are presented in \ref{append:box}. We have 
obtained explicit expressions for all box matrix elements with $L\leq 6$,
total spin $S\leq 2$, and total momentum $\Pvec=(0,0,0), (0,0,p)$,
as well as all box matrix elements with $L\leq 6$,
$S\leq \frac{3}{2}$, and $\Pvec=(0,p,p), (p,p,p)$, with $p>0$.
We have developed and tested software, written in C++, to evaluate these
box matrix elements.  This software is freely available\cite{gitavail} and
is described below.

Lastly, we need to express $\widetilde{K}$ in the new basis.
Given Eq.~(\ref{eq:Ksmooth}) and the orthonormality of the states
in both the $\ket{Jm_JLSa}$ basis and the block diagonal 
$\ket{\Lambda\lambda nJLSa}$ basis, one can show that
\beq
 \me{\Lambda'\lambda' n'J'L'S' a'}{\widetilde{K}}{\Lambda\lambda nJLS a}
= \delta_{\Lambda'\Lambda}\delta_{\lambda'\lambda}\delta_{n'n} \delta_{J'J}
\ {\cal K}^{(J)}_{L'S'a';\ LS a}(\Ecm),\quad (\eta=\eta'),
\label{eq:Kbdbasis}
\eeq
where $\eta=(-1)^L$ and $\eta'=(-1)^{L'}$.
If $\eta=-\eta^\prime$, the situation is much more complicated. However, in QCD,
we should never need such matrix elements.  Recall that invariance under parity 
and rotations implies that the $K$-matrix elements are nonzero only when
$\eta^{P\prime}_{1a}\eta^P_{1a}\eta^{P\prime}_{2a}\eta^P_{2a}(-1)^{L'+L}=1$.
All of the mesons we would ever want
to consider in the 2-particle states needed to relate finite-volume energies to
the $K$-matrix have negative parity: $\pi$, $\eta$, $\phi$, $K$, $D$, $B$, 
$\eta_b$, $\eta_c$, $J/\psi$, $\Upsilon$, and so on.  All of the baryons we 
would ever want to consider in 2-particle states have positive parity:
$N$, $\Delta$, $\Sigma$, $\Xi$, $\Omega$, $\Lambda$, $\Lambda_c$, $\Lambda_b$.
Thus, all meson-meson states of interest have $\eta^P_{1a}\eta^P_{2a}=1$,
and all meson-baryon states of interest have $\eta^P_{1a}\eta^P_{2a}=-1$,  
and in QCD, we should never need 2-to-2 $K$-matrix elements
between two-hadron states in which the products of the intrinsic parities are 
different.  In other field theories, it may occur that 
$\eta_{1a}^P\eta_{2a}^P=-\eta_{1a}^{P\prime}\eta_{2a}^{P\prime}$, in which case,
Eq.~(\ref{eq:Kbdbasis}) must be generalized.  

The box matrix is diagonal in total spin $S$ and in the compound index $a$.
However, the $\widetilde{K}$-matrix allows mixings between different spins
and channels, which means that the block structure of the box matrix is not 
the same as that of $1-B^{(\Pvec)}\widetilde{K}$ and $\widetilde{K}^{-1}-B^{(\Pvec)}$.
For a given $\Pvec,S,a$, the box matrix blocks can be labelled by $\Lambda_B,\lambda$, 
but the box matrix elements are independent of the irrep row $\lambda$.  The $K$-matrix
is diagonal in total angular momentum $J$ and is also independent of the irrep row
$\lambda$, as long as $\eta=\eta'$.  Given the independence of both matrices
on the irrep row, $\lambda$ is superfluous as a block identifier. Thus, for a given
$\Pvec$, we can label the quantization blocks of $1-B^{(\Pvec)}\widetilde{K}$ and 
$\widetilde{K}^{-1}-B^{(\Pvec)}$ in the $\vert\Lambda\lambda n JLSa\rangle$ basis 
solely by the irrep label $\Lambda$, where $\Lambda$ is the irrep associated
with the $K$-matrix.   

\section{Software overview}
\label{sec:software}

As the tables in \ref{append:box} show, the formulas for evaluating the box
matrix elements are cumbersome, and the evaluations of the RGL shifted zeta functions
are complicated and require care.  One goal of this work is to make software
available\cite{gitavail} that evaluates these quantities for use by practitioners.
A brief overview of the software is given here.

To evaluate $1-B^{(\Pvec)}\widetilde{K}$ or $\widetilde{K}^{-1}-B^{(\Pvec)}$ in
a given block for a particular $\Pvec$, the basis states involved in the block must first 
be determined.  To accomplish this, the individual spins $s_{1a}$ and $s_{2a}$ of 
particles 1 and 2 in each channel $a$ are needed, and a maximum orbital angular momentum 
$L_{\rm max}^{(a)}$ must be specified in each channel of particle species. Given the 
particle spins, the allowed total spins $S^{(a)}$ in each channel can be determined, 
and these are then combined with all allowed $L$ values up to $L_{\rm max}^{(a)}$ to 
produce the allowed $J$ values.  The software also checks to see if there are
any basis states $\alpha$ such that $\widetilde{K}_{\alpha\beta}$ are zero for
all states $\beta$, in which case, the state $\alpha$ is removed from the basis.
Once the basis of states is determined, 
Eqs.~(\ref{eq:Bbdform}) and (\ref{eq:Kbdbasis}) can then be used to evaluate 
$1-B^{(\Pvec)}\widetilde{K}$ or $\widetilde{K}^{-1}-B^{(\Pvec)}$ in this basis.  
Again, the irrep $\Lambda_B$ must be used for the box matrices contained in each 
$\Lambda$ quantization block for a given $\Pvec$. 

In our software, all masses and lengths are specified in terms of some reference
energy $m_{\rm ref}$.  The user must specify the vector of integers $\dvec$
specifying the total momentum, the little group irrep $\Lambda$, and the 
dimensionless quantity $m_{\rm ref}L$, as well as the number of channels.
For an anisotropic lattice of temporal spacing $a_t$ and spatial spacing $a_s$,
the aspect ratio $\xi=a_s/a_t$ must also be given.  In each channel, the 
user must give the dimensionless quantities $m_{1a}/m_{\rm ref},\ m_{2a}/m_{\rm ref}$ 
the particle spins $s_{1a}$ and $s_{2a}$, the product of intrinsic parities
$\eta^P_{1a}\eta^P_{2a}$, and a maximum orbital angular momentum $L_{\rm max}^{(a)}$.  With 
this information, the software sets up the basis of states.  When given a lab-frame 
energy $E/m_{\rm ref}$ or a center-of-momentum energy $\Ecm/m_{\rm ref}$, the
software evaluates the entire box matrix and the $\widetilde{K}$-matrix, then
returns the determinant of either $1-B^{(\Pvec)}\widetilde{K}$ or 
$\widetilde{K}^{-1}-B^{(\Pvec)}$.  To evaluate these quantities for different resamplings, 
the software can reset the masses and box lengths without rebuilding the basis.  The fit 
forms of Eqs.~(\ref{eq:Kfit1}) and (\ref{eq:Kfit2}) are used in the software, but these
can be easily modified if other forms are desired.

\section{Fitting}

Let $\kappa_j$, for $j=1,\dots,N_K$, denote the parameters that appear in the matrix 
elements of either the $\widetilde{K}$-matrix or its inverse $\widetilde{K}^{-1}$.  
Once a set of energies for a variety of two-particle interacting
states is determined, the primary goal is then to determine the best-fit estimates of the
$\kappa_j$ parameters using the quantization determinant, as well as to determine 
the uncertainties in these estimates. In this section, we describe two methods to 
achieve this.  

To set the stage, we first summarize the fitting procedure commonly used in 
lattice QCD. For an observable $O$, let ${\cal E}(O)$ denote a Monte Carlo estimate 
of the observable obtained using the entire Markov-chain ensemble of gauge 
configurations, and let ${\cal E}^{(r)}_k(O)$ denote an estimate of $O$ from the $k$-th 
resampling of a resampling scheme $r$.  The two most common resampling 
schemes are the jackknife $r=J$ and the bootstrap $r=B$.  If autocorrelations are small,
the covariance of the estimates of two observables $O_i$ and $O_j$ can be
estimated from their resampling estimates using 
\beqy
{\rm cov}(O_i,O_j) &\approx& {\cal N}^{(r)}\sum_{k=1}^{N_r} 
\Bigl({\cal E}^{(r)}_k(O_i)-\langle {\cal E}^{(r)}(O_i)\rangle\Bigr)
 \Bigl({\cal E}^{(r)}_k(O_j)-\langle {\cal E}^{(r)}(O_j)\rangle \Bigr),
\label{eq:covest}\\
   \langle{\cal E}^{(r)}(O_i)\rangle&=& \frac{1}{N_r}
 \sum_{k=1}^{N_r}{\cal E}^{(r)}_k(O_i),
\eeqy
where $N_r$ is the number of resamplings and the factor ${\cal N}^{(r)}$ depends 
on the resampling scheme. For the jackknife and bootstrap methods, it is given by
\beq
   {\cal N}^{(J)}= \frac{(N_J-1)}{N_J} ,\qquad {\cal N}^{(B)}=\frac{1}{N_B-1}.
\eeq
It often occurs that a set of observables is believed to be reasonably well described 
by a set of model functions containing unknown parameters.  In such cases, the goal 
is usually to find best fit estimates of these parameters.  Arrange the observables 
into the components of a vector $\Rvec$ and the fit parameters into a vector 
$\bm{\alpha}$.  Denote the set of model functions by the vector 
$\Mvec(\bm{\alpha},\Rvec)$ which depend on the parameters and which might depend on 
the observables themselves.  The $i$-th component of $\Mvec(\bm{\alpha},\Rvec)$ gives 
the model prediction for the observable corresponding to the $i$-th component of 
$\Rvec$.  In lattice QCD, we generally determine the best fit estimates of the 
$\bm{\alpha}$ parameters as the values which minimize a correlated$-\chi^2$ of 
residuals given by
\beq
    \chi^2 = {\cal E}(r_i)\ \sigma_{ij}^{-1}\ {\cal E}(r_j),
   \label{eq:chisqfit}
\eeq
where the vector of residuals is defined by
$\rvec(\Rvec,\bm{\alpha})=\Rvec-\Mvec(\bm{\alpha},\Rvec)$ and
$\sigma_{ij}={\rm cov}(r_i,r_j)$ is the covariance matrix of the residuals.
Since the observables are usually obtained using the same ensemble of gauge field
configurations, the residuals in Eq.~(\ref{eq:chisqfit}) are not statistically
independent so the presence of the covariance matrix in the likelihood function
is very important.

Usually, the minimization of $\chi^2$ with respect to the parameters $\bm{\alpha}$
is accomplished using computer software, such as \textsc{Minuit2} \cite{minuit2}.  If the model 
estimates of any of the observables depend on any of the other observables, then the 
covariance matrix must be recomputed using Eq.~(\ref{eq:covest}) and inverted each 
time the parameters $\bm{\alpha}$ are changed during the minimization process, making 
for a rather laborious minimization. A significant simplification occurs if the model
estimates are completely independent of the observables.  In this case, 
${\rm cov}(r_i,r_j)={\rm cov}(R_i,R_j)$, which needs to be computed and inverted only 
once at the start of the minimization.
The statistical uncertainties in the best-fit parameter estimates can usually be
obtained from the minimization software, which typically provides the covariances
of the parameter estimates under certain assumptions.  An alternative approach to 
determining these covariances is to perform the minimization of
\beq
    \chi^{2}_k = {\cal E}^{(r)}_k(r_i)\ \sigma_{ij}^{-1}\ {\cal E}^{(r)}_k(r_j),
   \label{eq:chisqfitk}
\eeq
for each resampling $k$ and obtain the covariance of the fit parameter estimates
${\rm cov}(\alpha_k,\alpha_l)$ using Eq.~(\ref{eq:covest}). 

Best fit estimates of the $\widetilde{K}$-matrix parameters can be improved
by utilizing results from multiple Markov-chain ensembles and lattices.  One
approach to performing such fits is to minimize the $\chi^2$ of 
Eq.~(\ref{eq:chisqfit}) taking the elements of $\sigma_{ij}$ to be zero between 
the estimates from different ensembles, then obtain the covariances of the best 
fit parameter estimates from the minimization software.  An alternative approach
is to ensure that $N_r$ is the same for all ensembles, then use the resamplings
of all ensembles in the $\chi^2$ of Eq.~(\ref{eq:chisqfit}) with the covariance
matrix estimated using Eq.~(\ref{eq:covest}).  Given the statistical independence
of the different ensembles, Eq.~(\ref{eq:covest}) naturally yields covariances 
between observable estimates from different ensembles which are very nearly zero. 

Again, the primary goal is to determine the best-fit estimates of the $\kappa_j$ 
parameters appearing in $\widetilde{K}$ or $\widetilde{K}^{-1}$ from the quantization
condition, as well as to determine the uncertainties in these estimates.  Having made 
the above introductory comments, we now describe two methods to achieve this.  

\subsection{Spectrum method}

For each $\Pvec$ and irrep $\Lambda$, one obtains as many lab frame
two-interacting-particle energies $E_k$ as possible, staying below the thresholds
for three or more particles.  For the observations $R_i$ in the fit, an 
obvious choice would be to
include the lab-frame energies $E_k$ or the center-of-momentum energies $\Ecmk$.  Here,
we choose the $\Ecmk$ energies. The quantization condition with the chosen 
functional forms of $\widetilde{K}$ or $\widetilde{K}^{-1}$ then provides the model 
predictions of the observations.  This involves scanning the quantization determinant 
in $\Ecm$ to find the values that result in the determinant having zero value.
Evaluating the determinant requires evaluating the box matrix elements, which
requires knowing $\svec_a, u_a^2$ for each channel $a$.  To determine $\svec_a, u^2_a$, 
one needs to know the masses $m_{1a}, m_{2a}$ in each decay channel, the
spatial lattice volume $L^3$, and the lattice aspect ratio $\xi=a_s/a_t$ if an
anisotropic lattice is used.  Unfortunately, these quantities must be obtained from 
the Monte Carlo simulations, and hence, are observations.  This poses the problem 
that the predictions cannot be obtained solely from the parameters of the model, 
independent of the observations.

A simple way around this problem is to include the masses $m_{1a}, m_{2a}$ in 
each decay channel, the spatial lattice volume $L^3$, and the lattice aspect 
ratio $\xi=a_s/a_t$ as both observations and model parameters.
In addition to the energy observations $\Ecmk^{({\rm obs})}$, one also includes 
$ m_{j}^{(\rm obs)},\ \ L^{(\rm obs)},\ \xi^{(\rm obs)},$
to the set of observations $R_i$, where $j=1,\dots,N_p$ and $N_p$ is the
number of different particle species in all of the decay channels.
At the same time, one introduces model parameters $m_{j}^{(\rm model)}, 
L^{(\rm model)}, \xi^{(\rm model)}$.  Now, in scanning for the zeros of the
determinant, one varies $\Ecm$, evaluating $\svec_a, u^2_a$, and hence
the box matrix elements, using the model parameters $m_{j}^{(\rm model)}, 
L^{(\rm model)}, \xi^{(\rm model)}$.  In doing this simple trick, the model 
predictions are independent of the observations. This procedure is somewhat 
in the spirit of introducing Lagrange multipliers in a minimization. 

In summary, the observations in the $\chi^2$ minimization in this first 
method are
\beq
   \mbox{Observations $R_i$:}\quad  \{ E_{{\rm cm},k}^{({\rm obs})},
  \ m_{j}^{(\rm obs)},\ L^{(\rm obs)},\ \xi^{(\rm obs)}\ \},
\eeq
for $k=1,\dots,N_E$ and $j=1,\dots,N_p$.
If there are $N_p$ particle species in all of the decay channels and $N_E$ energies
found, then there are $N_{\rm obs}=2+N_p+N_E$ observations on an anisotropic
lattice. Improved results can be 
obtained by increasing $N_E$ by using several different  $\Pvec, \Lambda$ blocks. 
The model parameters are
\beq
   \mbox{Model fit parameters $\alpha_k$:}\quad \{\ \kappa_i, \ m_{j}^{(\rm model)},
  \ L^{(\rm model)},\ \xi^{(\rm model)}\ \},
\label{eq:model1params}
\eeq
for $i=1,\dots,N_K$ and $j=1,\dots,N_p$, where $N_K$ is the total number
of parameters in the $\widetilde{K}$-matrix elements.  The total number of fit
parameters is $N_{\rm param}=2+N_p+N_K$. For an isotropic
lattice, the anisotropy observable and fit parameter can be omitted.  

Evaluating the predictions $M_i(\bm{\alpha})$ of the model for the $N_{\rm obs}$
observations is done as follows. The parameters $m_{j}^{(\rm model)},
\ L^{(\rm model)},\ \xi^{(\rm model)}$ themselves 
give the predictions for the observations $m_{j}^{(\rm obs)},
\ L^{(\rm obs)},\ \xi^{(\rm obs)}$.  The model predictions corresponding to the 
$E_{{\rm cm},k}^{({\rm obs})}$ observations are not so easily done.
One needs to scan the quantization determinant in $\Ecm$ to find which values
yield a zero value.  For a given $\Ecm$, one uses the parameters
$\kappa_j$ to evaluate the $\widetilde{K}$-matrix or its inverse, and determines the
box matrix elements in terms of the RGL zeta functions using the parameters 
$m_{j}^{(\rm model)},\ L^{(\rm model)},
\ \xi^{(\rm model)}$ to determine $\svec_a, u^2_a$.  This is a rather
onerous task.  Computing the determinant for a given $\Ecm$ is quite
complicated, and this must be done many times in order to bracket and then
numerically find all of the needed zeros of the determinant using
bisection or a Newton-Raphson type algorithm.  One must then match each root
found with the appropriate observed $\Ecm$.  Let $E_{{\rm cm},k}^{({\rm model})}$
denote each energy root found.  In summary, the residuals in this method are
\beq
    r_k = \left\{\begin{array}{ll} 
          E_{{\rm cm},k}^{({\rm obs})}-E_{{\rm cm},k}^{({\rm model})}, & (k=1,\dots,N_E),\\
          m_{k'}^{(\rm obs)}-m_{k'}^{(\rm model)}, & (k=k'+N_E,\ k'=1,\dots,N_p), \\
          L^{({\rm obs})}-L^{({\rm model})}, & (k=N_E+N_p+1),\\
          \xi^{({\rm obs})}-\xi^{({\rm model})},& (k=N_E+N_p+2).
   \end{array}\right.
\eeq
We emphasize that, in this method, the $E_{{\rm cm},k}^{({\rm model})}$ are
very difficult quantities to compute using the model parameters in 
Eq.~(\ref{eq:model1params}).  This method (without the model-parameter trick
and without properly treating all covariances) has been used in
Refs.~\cite{Dudek:2016cru,Wilson:2015dqa,Wilson:2014cna,Guo:2016zos}.

\subsection{Determinant residual method}

The difficulty in calculating the model predictions in the first method leads
us to seek other simpler methods.  In this second method, we introduce the
quantization determinant itself as a residual.  In the determinant, we use
the observed box matrix elements, which requires the observed energies and the 
observed values for the particle masses, lattice size, and anisotropy.  

Expressing the quantization condition in terms of a vanishing determinant
is just a convenient way of stating that one eigenvalue becomes zero.
The determinant itself is not a good quantity to use as an observable since
it can become very large in magnitude for larger matrices. Instead of the 
determinant, we express the quantization condition using the following
function of matrix $A$, having real determinant, and scalar $\mu\neq 0$:
\beq
   \Omega(\mu,A)\equiv \frac{\det(A)}{\det[(\mu^2+AA^\dagger)^{1/2}]}.
\eeq
When one of the eigenvalues of $A$ is zero, this function is also zero. 
This function can be evaluated as a product of terms, one for each
eigenvalue of $A$. For eigenvalues of $A$ which are much smaller in magnitude than 
$\vert\mu\vert$, the associated term in the product tends towards the 
eigenvalue itself, divided by $\vert\mu\vert$.  However, the key feature 
of this function is that for eigenvalues which are much larger than 
$\vert\mu\vert$, the associated term in the product goes to $e^{i\theta}$ 
for real $\theta$.  This function replaces the large unimportant 
eigenvalues with unimodular quantities so that the function does not grow 
with increasing matrix size.  This is a much better behaved function, 
bounded between -1 and 1 when the determinant is real, which still reproduces 
the quantization condition.  The constant $\mu$ can be chosen to optimize ease 
of numerical root finding or $\chi^2$ minimization.

In this method, the model fit parameters are just the $\kappa_i$ parameters,
and the residuals are chosen to be
\beq
    r_k = 
          \Omega\Bigl(\mu, 1-  B^{(\Pvec)}(E_{{\rm cm},k}^{({\rm obs})})
    \ \widetilde{K}(E_{{\rm cm},k}^{({\rm obs})})\Bigr), 
 \qquad  (k=1,\dots,N_E),
\eeq
or the matrix $\widetilde{K}(E_{{\rm cm},k}^{({\rm obs})})^{-1}-
B^{(\Pvec)}(E_{{\rm cm},k}^{({\rm obs})})$ could be used in the $\Omega$ function.
Clearly, the model predictions in this method are dependent on the observations 
themselves, so the covariance of the residual estimates must be recomputed and 
inverted by Cholesky decomposition throughout the minimization as the $\kappa_j$ 
parameters are adjusted.  However, this is still much simpler than the root 
finding required in the spectrum method.  

An advantage of this method is that 
the complicated RGL zeta functions only need to be computed for the box matrix 
elements as observables; they do not need to be recomputed as model
parameters are changed.  Since we cannot completely remove the dependence
of the model predictions on the observables in this method, there is no
advantage in introducing model parameters for the energies, particle masses, 
and the lattice anisotropy.  Hence, we do not need to recompute the box
matrix elements as the model parameters are adjusted in the $\chi^2$ 
minimization. The model predictions involve only the $\kappa_j$ parameters
and the observed energies, particle masses, and anisotropy.

Treating the box matrix elements as observables enables a natural interface between the 
lattice calculation and phenomenology.  If the box matrix elements and center-of-momentum
energies are calculated, then together with the covariances, they contain all the 
information required to extract the scattering amplitudes. Non-lattice practitioners can
use them without, for example, implementing the RGL zeta functions. These quantities
can act as a bridge between lattice QCD computations and phenomenological applications. 

\section{Tests of fitting procedures}

\begin{table}[p]
\caption[Fitting]{First tests of the determinant residual method applied to the
interacting $\pi\pi$ energies in the $I=1$ nonstrange channel described in
Ref.~\cite{Bulava:2016mks}.  These energies were obtained on a $32^3\times 256$
anisotropic lattice with $m_\pi\approx 240$~MeV.  In Ref.~\cite{Bulava:2016mks},
the number of energy levels used was $N_E=19$.  The fits below using $N_E=20$
include an additional energy from a $B_{1}^+\ d^2=1$ irrep in
which the leading partial wave is $L=3$. Fits which used $\det(\widetilde{K}^{-1}-B)$ 
as residuals are indicated by $-$ in the first column.  Fits with a value of
$\mu$ in the first column used $\Omega(\mu,\widetilde{K}^{-1}-B)$ as the residuals.
Fits in which higher partial waves were excluded are indicated by $-$
in the fifth and sixths columns corresponding to the scattering lengths of 
such waves.
\label{tab:tests}}
\begin{center}
\begin{tabular}{|ccccccl|}
\hline
$\mu$ &  $N_E$ & $m_{\rho}/m_{\pi}$  & $g$ &  $m_{\pi}^{7}a_3$    
  & $m_{\pi}^{11} a_5 $ & $\chi^2/\mathrm{dof}$ \\ \hline
\hline
$-$  & 19 &   3.352(24)   & 5.99(26)  & $-$ & $-$ & 1.04 \\ 
$-$  & 20 &   3.348(30)   & 5.95(28)  & -0.0010(64) & $-$ & 1.03 \\ 
$-$  & 19 &   3.35(11)   & 5.9(1.3)  & -0.0010(23) & $-$ & 1.10 \\ 
\hline
1   &  19 &  3.345(12)   & 6.02(18)  & $-$ & $-$ & 1.70 \\ 
1   &  20 &  3.345(13)   & 6.02(22)  & -0.00000186(16) & $-$ & 1.70 \\ 
1   &  19 &  3.342(22)   & 5.99(21)  & -0.0009(65) & $-$ & 1.79 \\ 
1   &  20 &  3.338(13)   & 5.91(17)  & 0.0001(12) & -0.00016(11) & 1.75 \\ 
\hline
2   &  19 &  3.351(18)   & 6.06(20)  & $-$ & $-$ & 1.41 \\ 
2   &  20 &  3.348(18)   & 6.04(20)  & -0.0010(12) & $-$ & 1.41 \\ 
2   &  19 &  3.348(19)   & 6.03(21)  & -0.0009(13) & $-$ & 1.48 \\ 
2   &  20 &  3.341(20)   & 5.91(22)  & 0.0001(15) & -0.00020(14) & 1.43 \\ 
\hline
4   &  19 &  3.353(22)   & 6.05(23)  & $-$ & $-$ & 1.21 \\ 
4   &  20 &  3.353(22)   & 6.06(23)  & -0.00017(10) & $-$ & 1.25 \\ 
4   &  19 &  3.391(33)   & 6.23(26)  & 0.0132(78) & $-$ & 1.70 \\ 
4   &  20 &  3.345(24)   & 5.92(26)  & 0.0001(17) & -0.00018(19) & 1.26 \\ 
\hline
8   &  19 &  3.352(23)   & 6.01(24)  & $-$ & $-$ & 1.10 \\ 
8   &  20 &  3.351(24)   & 6.01(25)  & -0.00041(64) & $-$ & 1.11 \\ 
8   &  19 &  3.351(29)   & 6.00(25)  & -0.0005(63) & $-$ & 1.17 \\ 
8   &  20 &  3.348(26)   & 5.96(28)  & -0.0001(12) & -0.00010(24) & 1.18 \\ 
\hline
10   & 19 &   3.352(23)   & 6.01(24)  & $-$ & $-$ & 1.08 \\ 
10   & 20 &   3.351(24)   & 6.02(25)  & -0.00008(26) & $-$ & 1.16 \\ 
10   & 19 &   3.350(27)   & 5.99(27)  & -0.0005(40) & $-$ & 1.15 \\ 
10   & 20 &   3.349(26)   & 5.97(27)  & -0.0002(11) & -0.00007(25) & 1.16 \\ 
\hline
12   &  19 &  3.352(23)   & 6.00(24)  & $-$ & $-$ & 1.07 \\ 
12   &  20 &  3.352(24)   & 6.00(26)  & -0.00000146(92) & $-$ & 1.07 \\ 
12   &  19 &  3.350(27)   & 5.99(27)  & -0.0005(41) & $-$ & 1.14 \\ 
12   &  20 &  3.349(25)   & 5.97(27)  & -0.00021(100) & -0.00006(24) & 1.15 \\ 
\hline
\end{tabular}
\end{center}
\end{table}

As first tests, we applied the determinant residual method to the interacting $\pi\pi$ 
energies in the $I=1$ nonstrange sector in irreps relevant for extracting the
$P$-wave amplitude.  The operators used and the energies obtained are 
described in Ref.~\cite{Bulava:2016mks}. These energies were obtained on a $32^3\times 256$
anisotropic lattice with $m_\pi\approx 240$~MeV.  
Defining
\beq
   k_0=\frac{2\pi}{m_\pi L},
\eeq
the fit forms we used are
\beqy
(\widetilde{K}^{-1})_{11} &=& \frac{6\pi\Ecm}{k_0^3m_{\pi} g^2}
\left(\frac{m_{\rho}^2}{m_{\pi}^2} - \frac{\Ecm^2}{m_{\pi}^2} \right), \\
(\widetilde{K}^{-1})_{33} &=& \frac{1}{k_0^7 m_{\pi}^7 a_3},\\
(\widetilde{K}^{-1})_{55} &=& \frac{1}{k_0^{11} m_{\pi}^{11} a_5}.
\eeqy

Our results are summarized in Table~\ref{tab:tests}. In Ref.~\cite{Bulava:2016mks},
the number of energy levels used was $N_E=19$.  In one half of our fits, we 
used just these $N_E=19$ levels.  In the other half of our fits, we also included 
an elastic energy from an additional $B_{1}^+\ d^2=1$ irrep in which the leading 
partial wave is $L=3$, as was done in Ref.~\cite{Wilson:2015dqa}.  These
fits are presented in the rows in Table~\ref{tab:tests} in which $N_E=20$.
Fits in which higher partial waves were excluded are indicated by $-$
in the fifth and sixths columns of Table~\ref{tab:tests}, corresponding to the 
scattering lengths of such waves.

In the first three rows of Table~\ref{tab:tests}, results are shown using
$\det(\widetilde{K}^{-1}-B)$ as the residuals.  The $\Omega$ function was not
used in these three fits.  Using the determinant alone, we were not able to perform 
the minimization including all $L=1,3,5$ partial waves.  Also, the errors on
the fit parameters using the determinant alone and using just the $N_E=19$
levels were found to be dramatically larger.  In all subsequent fits, we
utilized $\Omega(\mu,\widetilde{K}^{-1}-B)$ as the residuals.  Using the
$\Omega$ function, we were able to find the minimum of the $\chi^2$ function much
more easily.  For $\mu=1$, we found that the minimum $\chi^2/{\rm dof}$ values were
uncomfortably large.  This was remedied by increasing $\mu$ to a value
around $\mu=8$ or larger.

\begin{figure}
\begin{center}
\includegraphics[width=3.6in]{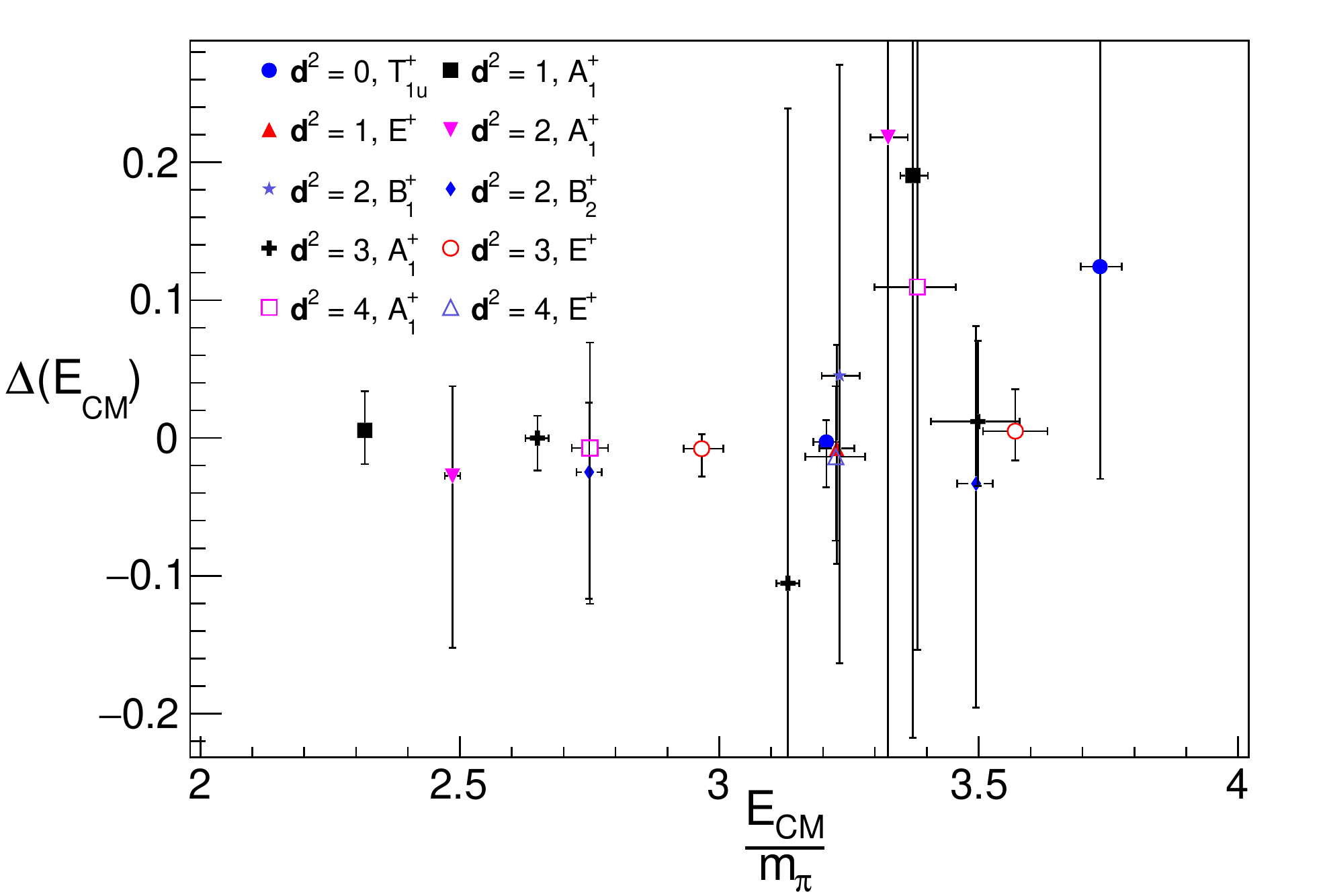}
\end{center}
\caption[cap1]{
The quantity $\Delta$, defined by Eq.~(\ref{eq:deltadef}) for each energy
where $L=1$ contributes, shows the
negligible effects of the higher $L=3$ and $L=5$ partial waves in the 
analysis of the $\rho$-decay to $\pi\pi$ using finite-volume energies.
\label{fig:boxdelta}}
\end{figure}

The most important thing that the test fits in Table~\ref{tab:tests} demonstrate 
is that the effects of higher partial waves can be taken into account using the 
determinant residual method.  Also, our results show that the phase shifts from the 
$L=3$ and $L=5$ waves are negligible in this energy range, justifying our neglect 
of these waves in Ref.~\cite{Bulava:2016mks}.  This is consistent with a 
phenomenological determination of $m_\pi^7a_3=5.65(21)\times 10^{-5}$ taken from 
Ref.~\cite{GarciaMartin:2011cn}.  The table also shows that the contribution
from the $L=3$ wave can be reliably estimated without using the additional
$B_{1}^+\ d^2=1$ irrep. To further illustrate the effect of the higher partial 
waves, we define a quantity $\Delta$ for each energy where $L=1$ contributes by
\beq
\widetilde{K}^{-1}_{11} = B^{(\Pvec)}_{11}( 1 + \Delta).
\label{eq:deltadef}
\eeq
This quantity is shown in Fig.~\ref{fig:boxdelta} for the $\mu=10,\ N_E=20$ 
fit including the three $L=1,3,5$ partial waves.  One sees that
$\Delta$ is consistent with zero throughout this energy range, again justifying
our neglect of the $L=3$ and $L=5$ waves in Ref.~\cite{Bulava:2016mks}.

In the future, we plan to utilize both the spectrum and residual determinant
methods in the analysis of meson-meson and meson-baryon systems
involving multiple channels.  Studies involving the $K^\ast(892)$ and
$a_0(980)$ should appear soon.  Various baryon resonances are also being
investigated.

\section{Conclusion}

The purpose of this paper was to provide explicit formulas, software, and new fitting
implementations for carrying out two-particle scattering studies in lattice QCD.
We introduced a so-called ``box matrix'' $B$ which describes how the partial
waves fit into the cubic finite volumes of lattice QCD simulations.  The
quantization condition was expressed in terms of this Hermitian matrix $B$ and
the real, symmetric scattering $K$-matrix.  The effective range expansion was
used to introduce an analytic matrix $\widetilde{K}^{-1}$.  We obtained explicit
expressions for the box matrix elements for several spins and center-of-momentum 
orbital angular momenta up to $L=6$ with total momentum of the form 
$\Pvec=(0,0,0), (0,0,p), (0,p,p), (p,p,p)$ for $p>0$.  More importantly,
the software for evaluating all of these box matrix elements was announced and
overviewed.  Lastly, we discussed two fitting strategies for estimating the
parameters used to approximate the $\widetilde{K}$-matrix.    First tests 
involving $\rho$-meson decay to two pions included the $L=3$ and $L=5$ partial 
waves, and the contributions from these higher waves were found to be negligible 
in the elastic energy range.

\vspace{4mm}
\noindent
{\bf Acknowledgements}: 
This work was supported by the U.S.~National Science Foundation 
under award PHY-1613449.  Computing resources were provided by
the Extreme Science and Engineering Discovery Environment (XSEDE)
under grant number TG-MCA07S017.  XSEDE is supported by National 
Science Foundation grant number ACI-1548562. We acknowledge helpful 
conversations with Raul Briceno and Steve Sharpe.  We dedicate this
work to the late Keisuke Jimmy Juge, our longtime collaborator and 
good friend.

\appendix
\section{Symmetries of the RGL shifted zeta functions}
\label{appendix:zeta}
\setcounter{table}{0}

In this appendix, we summarize the various symmetries of the RGL shifted zeta
functions, and some of the relations between them.
Using Eq.~(\ref{eq:zaccfinal2}), it is straightforward to show the following
relations and symmetries.  For any rotation $R$, one finds
\beqy
   {\cal Z}_{lm}(R\svec,\gamma,u^2)&=&\sum_{m'}
  {\cal Z}_{lm'}(\svec,\gamma,u^2)\  D^{(l)\ast}_{mm'}(R),
\label{eq:Zrotate1}\\
   {\cal Z}_{lm}(R^{-1}\svec,\gamma,u^2)&=&\sum_{m'}
  {\cal Z}_{lm'}(\svec,\gamma,u^2)\  D^{(l)}_{m'm}(R).
\label{eq:Zrotate}
\eeqy
Eq.~(\ref{eq:Zrotate1}) also applies for spatial inversion $R=I_s$ with
$I_s\svec = -\svec$ and
\beq
    D_{m'm}^{(l)}(I_s)=(-1)^l\ \delta_{m'm}.
\label{eq:Dparity}
\eeq
If the group element $R$ is in the little group of $\svec$, meaning, that
if $R\svec=\svec$, then $R^{-1}$ is also in the little group and
$R^{-1}\svec=\svec$.  We can use this fact and Eq.~(\ref{eq:Zrotate})
to find relationships between the different RGL shifted zeta functions
of the same $l$ but different $m$.  Applying Eq.~(\ref{eq:Zrotate}) to
all elements of the little group shows that many of the ${\cal Z}_{lm}$
are zero and that many of the nonzero ${\cal Z}_{lm}$ can be
expressed in terms of other RGL shifted zeta functions of the same $l$.  

Under $m\rightarrow -m$,  we have
\beq
     {\cal Z}_{l,-m}(\svec,\gamma,u^2)=(-1)^m
\ {\cal Z}_{lm}^\ast(\svec,\gamma,u^2).
\label{eq:Znegm}
\eeq
Under interchange of the two particle masses $m_1\leftrightarrow m_2$, one finds
\beq
 {\cal Z}_{lm}(\svec^{(m_1,m_2)},\gamma,u^2)=(-1)^l
 \ {\cal Z}_{lm}(\svec^{(m_2,m_1)},\gamma,u^2).
\eeq
When $m_1=m_2$, then the RGL shifted zeta functions must be zero for odd $l$.
When $m_1\neq m_2$, then the RGL shifted zeta functions are no longer
zero for odd values of $l$.
One last property of the RGL shifted zeta functions which follows from 
Eq.~(\ref{eq:zaccfinal2}) is
\beq
   {\rm Im}{\cal Z}_{l0}(\svec,\gamma,u^2)=0.
\eeq

For $\svec=(0,0,0)$, the little group is the orthogonal group $O_h$ which 
includes parity.  Applying these relations and solving, one finds that
the only nonzero values for angular momentum quantum numbers
$l\leq 12,\ 0\leq m\leq l$ are those listed in Table~\ref{tab:zsymrest}.
When $\dvec=(0,0,1)$, the little group is $C_{4v}$.   
Applying all group elements in the little group, and solving the resulting
relations, we find that the only nonzero values for $l\leq 12,\ 0\leq m\leq l$
are those listed in Table~\ref{tab:zsymonaxis}.
When $\dvec=(0,1,1)$, the little group is $C_{2v}$.  We have applied all group elements 
in the little group and solved the resulting relations for $l\leq 12$.
Results for $l\leq 9,\ 0\leq m\leq l$ are listed in Table~\ref{tab:zsympd}.
When $\dvec=(1,1,1)$, the little group is $C_{3v}$.  Again, we have determined
all relations for $l\leq 12$, but only results for $l\leq 8,\ 0\leq m\leq l$ are 
presented in Table~\ref{tab:zsymcd}.

\begin{table}[p]
\caption{Nonzero elements of ${\cal Z}_{lm}(\svec,\gamma,u^2)$ with $\svec=(0,0,0)$
for $l\leq 12,\ 0\leq m\leq l$. Negative values of $m$ can be obtained using
${\cal Z}_{l,-m}(\svec,\gamma,u^2)=(-1)^m\ {\cal Z}_{lm}^\ast(\svec,\gamma,u^2)$.
\label{tab:zsymrest}}
\begin{center}
\begin{tabular}{|l|l|l|} \hline
$l$ & Nonzero elements & Dependent nonzero elements \\ \hline\hline
0 & ${\rm Re}{\cal Z}_{00}$ & \\ \hline
4 & ${\rm Re}{\cal Z}_{40}$ & ${\rm Re}{\cal Z}_{44}=\sqrt{\frac{5}{14}}{\rm Re}{\cal Z}_{40}$\\ \hline
6 & ${\rm Re}{\cal Z}_{60}$ & ${\rm Re}{\cal Z}_{64}=-\sqrt{\frac{7}{2}}{\rm Re}{\cal Z}_{60}$\\ \hline
8 & ${\rm Re}{\cal Z}_{80}$ & ${\rm Re}{\cal Z}_{84}=\frac{\sqrt{14}}{3\sqrt{11}}
                            {\rm Re}{\cal Z}_{80}$ \\
  &                       & ${\rm Re}{\cal Z}_{88}=\frac{\sqrt{65}}{3\sqrt{22}}{\rm Re}{\cal Z}_{80}$ \\ \hline
10 & ${\rm Re}{\cal Z}_{10,0}$ 
   & ${\rm Re}{\cal Z}_{10, 4}=-\frac{\sqrt{66}}{\sqrt{65}}{\rm Re}{\cal Z}_{10,0}$ \\
  && ${\rm Re}{\cal Z}_{10, 8}=-\frac{\sqrt{187}}{\sqrt{130}} {\rm Re}{\cal Z}_{10,0}$ \\ \hline
12 & ${\rm Re}{\cal Z}_{12,0},\ {\rm Re}{\cal Z}_{12,4}$ &
 ${\rm Re}{\cal Z}_{12, 8} = \frac{\sqrt{429}}{\sqrt{646}} {\rm Re}{\cal Z}_{12,0}
       - \frac{\sqrt{672}}{\sqrt{323}} {\rm Re}{\cal Z}_{12,4}$ \\
&& ${\rm Re}{\cal Z}_{12, 12} = \frac{\sqrt{1456}}{\sqrt{7429}} {\rm Re}{\cal Z}_{12,0} 
      + \frac{\sqrt{891}}{\sqrt{7429}} {\rm Re}{\cal Z}_{12,4}$ \\ \hline
\end{tabular}
\end{center}
\end{table}

\begin{table}[p]
\caption{Nonzero elements of ${\cal Z}_{lm}(\svec,\gamma,u^2)$ with $\svec=(0,0,s)$, $s>0$,
for $l\leq 12,\ 0\leq m\leq l$. Negative values of $m$ can be obtained using
${\cal Z}_{l,-m}(\svec,\gamma,u^2)=(-1)^m\ {\cal Z}_{lm}^\ast(\svec,\gamma,u^2)$.
If the two particle masses are equal $m_1=m_2$, then the elements below for odd $l$
are zero.
\label{tab:zsymonaxis}}
\begin{center}
\begin{tabular}{|l|l|} \hline
$l$ & Nonzero elements \\ \hline\hline
0 & ${\rm Re}{\cal Z}_{00}$ \\ \hline
1 & ${\rm Re}{\cal Z}_{10}$ \\ \hline
2 & ${\rm Re}{\cal Z}_{20}$ \\ \hline
3 & ${\rm Re}{\cal Z}_{30}$ \\ \hline
4 & ${\rm Re}{\cal Z}_{40},\ {\rm Re}{\cal Z}_{44}$ \\ \hline
5 & ${\rm Re}{\cal Z}_{50},\ {\rm Re}{\cal Z}_{54}$ \\ \hline
6 & ${\rm Re}{\cal Z}_{60},\ {\rm Re}{\cal Z}_{64}$ \\ \hline
7 & ${\rm Re}{\cal Z}_{70},\ {\rm Re}{\cal Z}_{74}$ \\ \hline
8 & ${\rm Re}{\cal Z}_{80},\ {\rm Re}{\cal Z}_{84},\quad
     {\rm Re}{\cal Z}_{88}$ \\ \hline
9 &  ${\rm Re}{\cal Z}_{90},\ {\rm Re}{\cal Z}_{94},\ {\rm Re}{\cal Z}_{98}$\\ \hline
10 & ${\rm Re}{\cal Z}_{10,0},\ {\rm Re}{\cal Z}_{10,4},\ {\rm Re}{\cal Z}_{10,8}$\\ \hline
11 & ${\rm Re}{\cal Z}_{11,0},\ {\rm Re}{\cal Z}_{11,4},\ {\rm Re}{\cal Z}_{11,8}$\\ \hline
12 & ${\rm Re}{\cal Z}_{12,0},\ {\rm Re}{\cal Z}_{12,4},\ {\rm Re}{\cal Z}_{12,8},
     \ {\rm Re}{\cal Z}_{12,12}$\\ \hline
\end{tabular}
\end{center}
\end{table}

\begin{table}[p]
\caption{Nonzero elements of ${\cal Z}_{lm}(\svec,\gamma,u^2)$ with $\svec=(0,s,s)$, $s>0$,
for $l\leq 9,\ 0\leq m\leq l$. Negative values of $m$ can be obtained using
${\cal Z}_{l,-m}(\svec,\gamma,u^2)=(-1)^m\ {\cal Z}_{lm}^\ast(\svec,\gamma,u^2)$.
If the two particle masses are equal $m_1=m_2$, then the elements below for odd $l$
are zero.  We have also determined these relations for $9<l\leq 12$, but these are not
listed here.
\label{tab:zsympd}}
\begin{center}
\begin{tabular}{|l|l|l|} \hline
$l$ & Nonzero elements & Dependent nonzero elements \\ \hline\hline
0 & ${\rm Re}{\cal Z}_{00}$ & \\ \hline
1 & ${\rm Re}{\cal Z}_{10}$ & ${\rm Im}{\cal Z}_{11}=-\frac{1}{\sqrt{2}}{\rm Re}{\cal Z}_{10}$\\ \hline
2 & ${\rm Re}{\cal Z}_{20},\  {\rm Im}{\cal Z}_{21}$ & ${\rm Re}{\cal Z}_{22}= -\frac{\sqrt{6}}{2}{\rm Re}{\cal Z}_{20}$\\ \hline
3 & ${\rm Re}{\cal Z}_{30},\  {\rm Re}{\cal Z}_{32}$ & 
${\rm Im}{\cal Z}_{31}=\frac{\sqrt{3}}{4}{\rm Re}{\cal Z}_{30}+\frac{\sqrt{10}}{4}{\rm Re}{\cal Z}_{32}$ \\ 
&&${\rm Im}{\cal Z}_{33}=\frac{\sqrt{5}}{4}{\rm Re}{\cal Z}_{30}-\frac{\sqrt{6}}{4}{\rm Re}{\cal Z}_{32}$ \\ \hline
4 & ${\rm Re}{\cal Z}_{40},\ {\rm Im}{\cal Z}_{41},$
& ${\rm Im}{\cal Z}_{43}=-\sqrt{7}{\rm Im}{\cal Z}_{41}$\\ 
& ${\rm Re}{\cal Z}_{42}$ &${\rm Re}{\cal Z}_{44}=\frac{\sqrt{5}}{\sqrt{14}}{\rm Re}{\cal Z}_{40}-\frac{2}{\sqrt{7}}{\rm Re}{\cal Z}_{42}$\\ \hline
5 & ${\rm Re}{\cal Z}_{50},\  {\rm Re}{\cal Z}_{52},$
&${\rm Im}{\cal Z}_{51}= -\frac{\sqrt{30}}{16}   {\rm Re}{\cal Z}_{50} -\frac{\sqrt{7}}{4}     {\rm Re}{\cal Z}_{52} - \frac{\sqrt{21}}{8}    {\rm Re}{\cal Z}_{54}$\\
& ${\rm Re}{\cal Z}_{54}$ &$ {\rm Im}{\cal Z}_{53}= -\frac{\sqrt{35}}{16}    {\rm Re}{\cal Z}_{50} -\frac{\sqrt{6}}{8}     {\rm Re}{\cal Z}_{52} + \frac{9\sqrt{2}}{16}    {\rm Re}{\cal Z}_{54}$\\
&&$ {\rm Im}{\cal Z}_{55}=-\frac{3\sqrt{7}}{16}   {\rm Re}{\cal Z}_{50} +\frac{\sqrt{30}}{8}     {\rm Re}{\cal Z}_{52} - \frac{\sqrt{10}}{16}    {\rm Re}{\cal Z}_{54}$\\ \hline
6 & ${\rm Re}{\cal Z}_{60},\ {\rm Im}{\cal Z}_{61},$
&${\rm Re}{\cal Z}_{64} =  - \frac{\sqrt{7}}{\sqrt{2}} {\rm Re}{\cal Z}_{60} - \frac{8\sqrt{2}}{\sqrt{15}}  {\rm Re}{\cal Z}_{62} $\\
& ${\rm Re}{\cal Z}_{62},\ {\rm Im}{\cal Z}_{63}$
&${\rm Im}{\cal Z}_{65} = \frac{\sqrt{6}}{\sqrt{11}} {\rm Im}{\cal Z}_{61} - \frac{\sqrt{15}}{\sqrt{11}}  {\rm Im}{\cal Z}_{63} $\\ 
&&${\rm Re}{\cal Z}_{66} =  \frac{\sqrt{11}}{\sqrt{5}}  {\rm Re}{\cal Z}_{62} $\\ \hline
7 & ${\rm Re}{\cal Z}_{70},\ {\rm Re}{\cal Z}_{72},$ 
& ${\rm Im}{\cal Z}_{71} = \frac{5 \sqrt{14}}{64}   {\rm Re}{\cal Z}_{70} \!+\! \frac{15 \sqrt{6}}{64}   {\rm Re}{\cal Z}_{72} \!+\! \frac{3 \sqrt{33}}{32}   {\rm Re}{\cal Z}_{74} \!+\! \frac{\sqrt{858}}{64}   {\rm Re}{\cal Z}_{76} $\\
& ${\rm Re}{\cal Z}_{74},\ {\rm Re}{\cal Z}_{76}$ 
& ${\rm Im}{\cal Z}_{73} = \frac{3 \sqrt{42}}{64}   {\rm Re}{\cal Z}_{70} \!+\! \frac{19 \sqrt{2}}{64}   {\rm Re}{\cal Z}_{72} \!-\! \frac{\sqrt{11}}{32}   {\rm Re}{\cal Z}_{74} \!-\! \frac{3 \sqrt{286}}{64}   {\rm Re}{\cal Z}_{76} $\\
&&${\rm Im}{\cal Z}_{75} = \frac{\sqrt{462}}{64 }  {\rm Re}{\cal Z}_{70} \!+\! \frac{\sqrt{22}}{64}   {\rm Re}{\cal Z}_{72} \!-\! \frac{25}{32}   {\rm Re}{\cal Z}_{74} \!+\! \frac{5 \sqrt{26}}{64}   {\rm Re}{\cal Z}_{76} $\\
&&${\rm Im}{\cal Z}_{77} = \frac{\sqrt{858}}{64}   {\rm Re}{\cal Z}_{70} \!-\! \frac{\sqrt{2002}}{64}   {\rm Re}{\cal Z}_{72} \!+\! \frac{\sqrt{91}}{32}   {\rm Re}{\cal Z}_{74} \!-\! \frac{\sqrt{14}}{64}   {\rm Re}{\cal Z}_{76} $\\ \hline
8 & ${\rm Re}{\cal Z}_{80},\ {\rm Im}{\cal Z}_{81},$ 
& ${\rm Im}{\cal Z}_{85} =  -\frac{2 \sqrt{77}}{\sqrt{13}}   {\rm Im}{\cal Z}_{81} - \frac{3 \sqrt{15}}{\sqrt{13}}   {\rm Im}{\cal Z}_{83} $\\
 & ${\rm Re}{\cal Z}_{82},\ {\rm Im}{\cal Z}_{83},$ 
& ${\rm Re}{\cal Z}_{86} = \frac{4 \sqrt{3}}{\sqrt{143}}   {\rm Re}{\cal Z}_{80} + \frac{\sqrt{15}}{\sqrt{1001}}   {\rm Re}{\cal Z}_{82} - \frac{6 \sqrt{6}}{\sqrt{91} }  {\rm Re}{\cal Z}_{84} $\\
 & ${\rm Re}{\cal Z}_{84}$ 
& ${\rm Im}{\cal Z}_{87} = \frac{ \sqrt{55}}{\sqrt{13}}   {\rm Im}{\cal Z}_{81} + \frac{2 \sqrt{21}}{\sqrt{13}}   {\rm Im}{\cal Z}_{83} $\\
&& ${\rm Re}{\cal Z}_{88} = \frac{3 \sqrt{5}}{\sqrt{286}}   {\rm Re}{\cal Z}_{80} - \frac{16 \sqrt{2}}{\sqrt{1001}}   {\rm Re}{\cal Z}_{82} + \frac{2 \sqrt{5}}{\sqrt{91}}   {\rm Re}{\cal Z}_{84} $\\ \hline
9  &  ${\rm Re}{\cal Z}_{90},\ {\rm Re}{\cal Z}_{92},$ 
 & ${\rm Im}{\cal Z}_{91} =  - \frac{\sqrt{2205}}{\sqrt{32768}}  {\rm Re}{\cal Z}_{90} - \frac{\sqrt{539}}{\sqrt{2048}}  {\rm Re}{\cal Z}_{92} - \frac{\sqrt{1001}}{\sqrt{4096}}  {\rm Re}{\cal Z}_{94}$\\
   &  ${\rm Re}{\cal Z}_{94},\ {\rm Re}{\cal Z}_{96},$ 
 & $\qquad\qquad  - \frac{\sqrt{429}}{\sqrt{2048}}  {\rm Re}{\cal Z}_{96} - \frac{\sqrt{2431}}{\sqrt{16384}}  {\rm Re}{\cal Z}_{98}$\\
   &  ${\rm Re}{\cal Z}_{98}$ 
 &  ${\rm Im}{\cal Z}_{93} =  - \frac{\sqrt{1155}}{\sqrt{16384}}  {\rm Re}{\cal Z}_{90} - \frac{\sqrt{189}}{\sqrt{1024}}  {\rm Re}{\cal Z}_{92} - \frac{\sqrt{39}}{\sqrt{2048}}  {\rm Re}{\cal Z}_{94}$\\
 & & $\qquad\qquad + \frac{\sqrt{91}}{\sqrt{1024}}  {\rm Re}{\cal Z}_{96} + \frac{\sqrt{4641}}{\sqrt{8192}}  {\rm Re}{\cal Z}_{98}$\\
 && ${\rm Im}{\cal Z}_{95} =  - \frac{\sqrt{1287}}{\sqrt{16384}}  {\rm Re}{\cal Z}_{90}  - \frac{\sqrt{65}}{\sqrt{1024}}  {\rm Re}{\cal Z}_{92} + \frac{\sqrt{315}}{\sqrt{2048}}  {\rm Re}{\cal Z}_{94}$\\ &&$\qquad\qquad+ \frac{\sqrt{375}}{\sqrt{1024}}  {\rm Re}{\cal Z}_{96} - \frac{\sqrt{2125}}{\sqrt{8192}}  {\rm Re}{\cal Z}_{98}$ \\
 &&${\rm Im}{\cal Z}_{97} =  - \frac{\sqrt{6435}}{\sqrt{65536}}  {\rm Re}{\cal Z}_{90} + \frac{\sqrt{13}}{\sqrt{4096}}  {\rm Re}{\cal Z}_{92} + \frac{\sqrt{3703}}{\sqrt{8192}}  {\rm Re}{\cal Z}_{94}$\\ &&$\qquad\qquad- \frac{\sqrt{1323}}{\sqrt{4096}}  {\rm Re}{\cal Z}_{96} + \frac{\sqrt{833}}{\sqrt{32768}}  {\rm Re}{\cal Z}_{98}$ \\
 && ${\rm Im}{\cal Z}_{99} =  - \frac{\sqrt{12155}}{\sqrt{65536}}  {\rm Re}{\cal Z}_{90} + \frac{\sqrt{1989}}{\sqrt{4096}}  {\rm Re}{\cal Z}_{92} - \frac{\sqrt{1071}}{\sqrt{8192}}  {\rm Re}{\cal Z}_{94}$\\ &&$\qquad\qquad+ \frac{\sqrt{51}}{\sqrt{4096}}  {\rm Re}{\cal Z}_{96} - \frac{\sqrt{9}}{\sqrt{32768}}  {\rm Re}{\cal Z}_{98}$\\ \hline
\end{tabular}
\end{center}
\end{table}

\begin{table}[p]
\caption{Nonzero elements of ${\cal Z}_{lm}(\svec,\gamma,u^2)$ with $\svec=(s,s,s)$, $s>0$,
for $l\leq 8,\ 0\leq m\leq l$. Negative values of $m$ can be obtained using
${\cal Z}_{l,-m}(\svec,\gamma,u^2)=(-1)^m\ {\cal Z}_{lm}^\ast(\svec,\gamma,u^2)$.
If the two particle masses are equal $m_1=m_2$, then the elements below for odd $l$
are zero.  We have also determined these relations for $8<l\leq 12$, but these are not
shown here.
\label{tab:zsymcd}}
\begin{center}
\begin{tabular}{|l|l|l|} \hline
$l$ & Nonzero elements & Dependent nonzero elements \\ \hline\hline
0 & ${\rm Re}{\cal Z}_{00}$ & \\ \hline
1 & ${\rm Re}{\cal Z}_{10}$
& ${\rm Re}{\cal Z}_{11}={\rm Im}{\cal Z}_{11}=-\frac{1}{\sqrt{2}}{\rm Re}{\cal Z}_{10}$\\ \hline
2 & ${\rm Re}{\cal Z}_{21}$ &
 ${\rm Im}{\cal Z}_{21} =  -{\rm Im}{\cal Z}_{22} ={\rm Re}{\cal Z}_{21}$ \\ \hline
3 & ${\rm Re}{\cal Z}_{30},\ {\rm Im}{\cal Z}_{32}$ &
${\rm Re}{\cal Z}_{31} = {\rm Im}{\cal Z}_{31} = \frac{\sqrt{3}}{4} {\rm Re}{\cal Z}_{30}$\\
&&${\rm Re}{\cal Z}_{33} = -{\rm Im}{\cal Z}_{33} = -\frac{\sqrt{5}}{4}{\rm Re}{\cal Z}_{30}$\\ \hline
4 & ${\rm Re}{\cal Z}_{40},\ {\rm Re}{\cal Z}_{41}$ &
${\rm Im}{\cal Z}_{42} =  2\sqrt{2}\, {\rm Re}{\cal Z}_{41}$\\
&&${\rm Re}{\cal Z}_{43} =  -{\rm Im}{\cal Z}_{43} = \sqrt{7}\, {\rm Re}{\cal Z}_{41}$\\
&&${\rm Im}{\cal Z}_{41} =  {\rm Re}{\cal Z}_{41}$\\
&&${\rm Re}{\cal Z}_{44} =  \frac{\sqrt{5}}{\sqrt{14}} {\rm Re}{\cal Z}_{40}$\\ \hline
5 & ${\rm Re}{\cal Z}_{50},\ {\rm Re}{\cal Z}_{51}$ &
${\rm Im}{\cal Z}_{51} = {\rm Re}{\cal Z}_{51}$\\
&&${\rm Re}{\cal Z}_{53} = - {\rm Im}{\cal Z}_{53} =  \frac{\sqrt{5}}{\sqrt{7}} {\rm Re}{\cal Z}_{50} + \frac{3\sqrt{3}}{\sqrt{14}} {\rm Re}{\cal Z}_{51}$\\
&&${\rm Re}{\cal Z}_{54} = - \frac{\sqrt{5}}{\sqrt{14}} {\rm Re}{\cal Z}_{50} -\frac{8}{\sqrt{21}} {\rm Re}{\cal Z}_{51}$\\
&&${\rm Re}{\cal Z}_{55} = {\rm Im}{\cal Z}_{55} = - \frac{1}{\sqrt{7}} {\rm Re}{\cal Z}_{50} + \frac{\sqrt{5}}{\sqrt{42}} {\rm Re}{\cal Z}_{51}$\\ \hline
6 & ${\rm Re}{\cal Z}_{60},\ {\rm Re}{\cal Z}_{61},$ &
 ${\rm Im}{\cal Z}_{61} = {\rm Re}{\cal Z}_{61}$\\
& ${\rm Re}{\cal Z}_{63}$ 
& ${\rm Im}{\cal Z}_{62} =  - \frac{\sqrt{10}}{4} {\rm Re}{\cal Z}_{61} - \frac{3}{2} {\rm Re}{\cal Z}_{63}$\\
&&${\rm Im}{\cal Z}_{63} =  - {\rm Re}{\cal Z}_{63}$\\
&&${\rm Re}{\cal Z}_{64} =  - \frac{\sqrt{7}}{\sqrt{2}} {\rm Re}{\cal Z}_{60}$\\
&&${\rm Re}{\cal Z}_{65} = {\rm Im}{\cal Z}_{65} =  \frac{\sqrt{6}}{\sqrt{11}} {\rm Re}{\cal Z}_{61} + \frac{\sqrt{15}}{\sqrt{11}} {\rm Re}{\cal Z}_{63} $\\
&&${\rm Im}{\cal Z}_{66} =  - \frac{9}{2\sqrt{22}} {\rm Re}{\cal Z}_{61} + \frac{\sqrt{5}}{2\sqrt{11}}{\rm Re}{\cal Z}_{63} $\\ \hline
7 & $ {\rm Re}{\cal Z}_{70},\ {\rm Re}{\cal Z}_{71}, $ &
 ${\rm Im}{\cal Z}_{71} =  {\rm Re}{\cal Z}_{71}$ \\
  & $ {\rm Im}{\cal Z}_{72} $ &
 ${\rm Re}{\cal Z}_{73} = -{\rm Im}{\cal Z}_{73} = - \frac{\sqrt{7}}{3\sqrt{6}}  {\rm Re}{\cal Z}_{70} + \frac{1}{3\sqrt{3}}  {\rm Re}{\cal Z}_{71}$\\
&& ${\rm Re}{\cal Z}_{74} =  - \frac{5\sqrt{7}}{3\sqrt{66}}  {\rm Re}{\cal Z}_{70} + \frac{32}{3\sqrt{33}}  {\rm Re}{\cal Z}_{71}   $\\
&& ${\rm Re}{\cal Z}_{75} = {\rm Im}{\cal Z}_{75} = \frac{7\sqrt{7}}{3\sqrt{66}}  {\rm Re}{\cal Z}_{70} - \frac{25}{3\sqrt{33}}  {\rm Re}{\cal Z}_{71} $\\
&& ${\rm Im}{\cal Z}_{76} =  \frac{\sqrt{11}}{\sqrt{13}}  {\rm Im}{\cal Z}_{72}$\\
&& ${\rm Re}{\cal Z}_{77} =  -{\rm Im}{\cal Z}_{77} =- \frac{\sqrt{26}}{3\sqrt{33}} {\rm Re}{\cal Z}_{70} - \frac{\sqrt{91}}{3\sqrt{33}}  {\rm Re}{\cal Z}_{71} $\\ \hline
8 & ${\rm Re}{\cal Z}_{80},\ {\rm Re}{\cal Z}_{81},$ &
  ${\rm Im}{\cal Z}_{81} =  {\rm Re}{\cal Z}_{81}$\\
 & ${\rm Re}{\cal Z}_{83}$ &
${\rm Im}{\cal Z}_{82} =  - \frac{\sqrt{35}}{\sqrt{2}}  {\rm Re}{\cal Z}_{81} + \frac{\sqrt{33}}{\sqrt{2}}  {\rm Re}{\cal Z}_{83}$\\
&&${\rm Im}{\cal Z}_{83} =  - {\rm Re}{\cal Z}_{83}$\\
&&${\rm Re}{\cal Z}_{84} =  \frac{\sqrt{14}}{3\sqrt{11}} {\rm Re}{\cal Z}_{80}$\\
&&${\rm Re}{\cal Z}_{85} = {\rm Im}{\cal Z}_{85} = - \frac{2\sqrt{77}}{\sqrt{13}}  {\rm Re}{\cal Z}_{81} + \frac{3\sqrt{15}}{\sqrt{13}}  {\rm Re}{\cal Z}_{83} $\\
&&${\rm Im}{\cal Z}_{86} =  \frac{3\sqrt{33}}{\sqrt{26}}  {\rm Re}{\cal Z}_{81} - \frac{\sqrt{35}}{\sqrt{26}}  {\rm Re}{\cal Z}_{83}$\\
&&${\rm Re}{\cal Z}_{87} =  -{\rm Im}{\cal Z}_{87} =- \frac{\sqrt{55}}{\sqrt{13}}  {\rm Re}{\cal Z}_{81} + \frac{2\sqrt{21}}{\sqrt{13}}  {\rm Re}{\cal Z}_{83} $\\
&&${\rm Re}{\cal Z}_{88} =  \frac{\sqrt{65}}{3\sqrt{22}}  {\rm Re}{\cal Z}_{80}$\\ \hline
\end{tabular}
\end{center}
\end{table}

\section{Box matrix elements in the block diagonal basis}
\label{append:box}
\setcounter{table}{0}

Expressions in terms of the RGL shifted zeta functions for a small 
selection of the box matrix elements 
$B^{(\Pvec\Lambda_B S a)}_{J'L'n';\ JLn}(E)$ which we have determined
are presented in Tables~\ref{tab:boxP0S0A}-\ref{tab:boxP22S1A}. 
These quantities depend on $a$ only through $\svec_a$ and $u_a$.  We have 
obtained explicit expressions for all box matrix elements with $L\leq 6$,
total spin $S\leq 2$, and total momentum $\Pvec=(0,0,0), (0,0,p)$,
as well as all box matrix elements with $L\leq 6$,
$S\leq \frac{3}{2}$, and $\Pvec=(0,p,p), (p,p,p)$, with $p>0$.
These are available from Ref.~\cite{gitavail}.  The software for evaluating these
box matrix elements was described in Sec.~\ref{sec:software} and is freely 
available\cite{gitavail}.

\begin{table}[p]
\caption{
Box matrix elements $B^{(\Pvec\Lambda_B S a)}_{J'L'n';\ JLn}(E)$
for various irreps with $\Pvec=0$ and total spin $S=0$. 
These quantities depend on $a$ only through $\svec_a$ and $u_a$.  
$R_{lm}$ is short hand
for $(\gamma \pi^{3/2}u_a^{l+1})^{-1}{\rm Re}\ {\cal Z}_{lm}(\svec_a,\gamma,u_a^2)$.
The Hermiticity of $B$ can be used to obtain other elements that are not shown.
The irrep $A_{1u}$ does not occur for $\Pvec=0,\ S=0$.
\label{tab:boxP0S0A}}
\begin{center}
\begin{tabular}{|ccc|ccc|l|}\hline
$J'$ & $L'$ & $n'$ & $J$ & $L$& $n$ & $\qquad\qquad u_a^{-(L'+L+1)}\ B$ \\ \hline 
\multicolumn{7}{c}{}\\[-5pt] 
\multicolumn{7}{c}{$\Lambda_B=A_{1g}$}\\ \hline
0 & 0 & 1 & 0 & 0 & 1 & $ R_{{00}}$\\
0 & 0 & 1 & 4 & 4 & 1 & $\frac{2\sqrt {21}}{7} R_{{40}}$\\
0 & 0 & 1 & 6 & 6 & 1 & $-2\sqrt {2} R_{{60}}$\\
4 & 4 & 1 & 4 & 4 & 1 & $  R_{{00}} +{\frac {108}{143}} R_{{40}} +{\frac {80\,\sqrt {13}}{143}} R_{{60}} +{\frac {560\,\sqrt {17}}{2431}} R_{{80}}$\\
4 & 4 & 1 & 6 & 6 & 1 & $ -{\frac {40\,\sqrt {546}}{1001}} R_{{40}} +{\frac {42\,\sqrt {42}}{187}} R_{{60}} -{\frac {224\,\sqrt {9282}}{46189}} R_{{80}} -{\frac {1008\,\sqrt {26}}{4199}} R_{{10,0}}$\\
6 & 6 & 1 & 6 & 6 & 1 & $  R_{{00}} -{\frac {126}{187}} R_{{40}} -{\frac {160\,\sqrt {13}}{3553}} R_{{60}} +{\frac {840\,\sqrt {17}}{3553}} R_{{80}} -{\frac {2016\,\sqrt {21}}{7429}} R_{{10,0}}$\\
 &&&&&& $\qquad  +{\frac {30492}{37145}} R_{{12,0}} -{\frac {1848\,\sqrt {1001}}{37145}} R_{{12,4}}$\\
\hline
\multicolumn{7}{c}{}\\[-5pt] 
\multicolumn{7}{c}{$\Lambda_B=A_{2g}$}\\ \hline
6 & 6 & 1 & 6 & 6 & 1 & $  R_{{00}} +{\frac {6}{17}} R_{{40}} -{\frac {160\,\sqrt {13}}{323}} R_{{60}} -{\frac {40\,\sqrt {17}}{323}} R_{{80}} -{\frac {2592\,\sqrt {21}}{7429}} R_{{10,0}}$\\
 &&&&&& $\qquad  +{\frac {1980}{7429}} R_{{12,0}} +{\frac {264\,\sqrt {1001}}{7429}} R_{{12,4}}$\\
\hline
\multicolumn{7}{c}{}\\[-5pt] 
\multicolumn{7}{c}{$\Lambda_B=A_{2u}$}\\ \hline
3 & 3 & 1 & 3 & 3 & 1 & $  R_{{00}} -{\frac {12}{11}} R_{{40}} +{\frac {80\,\sqrt {13}}{143}} R_{{60}}$\\
\hline
\multicolumn{7}{c}{}\\[-5pt] 
\multicolumn{7}{c}{$\Lambda_B=E_{g}$}\\ \hline
2 & 2 & 1 & 2 & 2 & 1 & $  R_{{00}} +\frac{6}{7} R_{{40}}$\\
2 & 2 & 1 & 4 & 4 & 1 & $ -{\frac {40\,\sqrt {3}}{77}} R_{{40}} -{\frac {30\,\sqrt {39}}{143}} R_{{60}}$\\
2 & 2 & 1 & 6 & 6 & 1 & $ {\frac {30\,\sqrt {910}}{1001}} R_{{40}} +{\frac {4\,\sqrt {70}}{55}} R_{{60}} +{\frac {8\,\sqrt {15470}}{1105}} R_{{80}}$\\
4 & 4 & 1 & 4 & 4 & 1 & $  R_{{00}} +{\frac {108}{1001}} R_{{40}} -{\frac {64\,\sqrt {13}}{143}} R_{{60}} +{\frac {392\,\sqrt {17}}{2431}} R_{{80}}$\\
4 & 4 & 1 & 6 & 6 & 1 & $ -{\frac {8\,\sqrt {2730}}{1001}} R_{{40}} -{\frac {18\,\sqrt {210}}{187}} R_{{60}} -{\frac {128\,\sqrt {46410}}{46189}} R_{{80}} $\\
 &&&&&& $\qquad-{\frac {1512\,\sqrt {130}}{20995}} R_{{10,0}}$\\
6 & 6 & 1 & 6 & 6 & 1 & $  R_{{00}} +{\frac {114}{187}} R_{{40}} +{\frac {480\,\sqrt {13}}{3553}} R_{{60}} +{\frac {280\,\sqrt {17}}{3553}} R_{{80}} +{\frac {1152\,\sqrt {21}}{7429}} R_{{10,0}}$\\
 &&&&&& $\qquad  +{\frac {30492}{37145}} R_{{12,0}} +{\frac {264\,\sqrt {1001}}{37145}} R_{{12,4}}$\\
\hline
\multicolumn{7}{c}{}\\[-5pt] 
\multicolumn{7}{c}{$\Lambda_B=E_{u}$}\\ \hline
5 & 5 & 1 & 5 & 5 & 1 & $  R_{{00}} -{\frac {6}{13}} R_{{40}} +{\frac {32\,\sqrt {13}}{221}} R_{{60}} -{\frac {672\,\sqrt {17}}{4199}} R_{{80}} +{\frac {1152\,\sqrt {21}}{4199}} R_{{10,0}}$\\
\hline
\multicolumn{7}{c}{}\\[-5pt]
\multicolumn{7}{c}{$\Lambda_B=T_{1g}$}\\ \hline
4 & 4 & 1 & 4 & 4 & 1 & $  R_{{00}} +{\frac {54}{143}} R_{{40}} -{\frac {4\,\sqrt {13}}{143}} R_{{60}} -{\frac {448\,\sqrt {17}}{2431}} R_{{80}}$\\
4 & 4 & 1 & 6 & 6 & 1 & $ -{\frac {12\,\sqrt {65}}{143}} R_{{40}} +{\frac {42\,\sqrt {5}}{187}} R_{{60}} +{\frac {112\,\sqrt {1105}}{46189}} R_{{80}} +{\frac {576\,\sqrt {1365}}{20995}} R_{{10,0}}$\\
6 & 6 & 1 & 6 & 6 & 1 & $  R_{{00}} -{\frac {96}{187}} R_{{40}} -{\frac {80\,\sqrt {13}}{3553}} R_{{60}} +{\frac {120\,\sqrt {17}}{3553}} R_{{80}} +{\frac {624\,\sqrt {21}}{7429}} R_{{10,0}}$\\
 &&&&&& $\qquad  -{\frac {26136}{37145}} R_{{12,0}} +{\frac {1584\,\sqrt {1001}}{37145}} R_{{12,4}}$\\
\hline
\end{tabular}
\end{center}
\end{table}

\begin{table}[p]
\caption{
Box matrix elements $B^{(\Pvec\Lambda_B S a)}_{J'L'n';\ JLn}(E)$
for various irreps with $\Pvec=0$ and total spin $S=0$.
These quantities depend on $a$ only through $\svec_a$ and $u_a$.  
$R_{lm}$ is short hand
for $(\gamma \pi^{3/2}u_a^{l+1})^{-1}{\rm Re}\ {\cal Z}_{lm}(\svec_a,\gamma,u_a^2)$.
The Hermiticity of $B$ can be
used to obtain other elements that are not shown.
\label{tab:boxP0S0B}}
\begin{center}
\begin{tabular}{|ccc|ccc|l|}\hline
$J'$ & $L'$ & $n'$ & $J$ & $L$& $n$ & $\qquad\qquad u_a^{-(L'+L+1)}\ B$ \\ \hline
\multicolumn{7}{c}{}\\[-5pt]
\multicolumn{7}{c}{$\Lambda_B=T_{1u}$}\\ \hline
1 & 1 & 1 & 1 & 1 & 1 & $ R_{{00}}$\\
1 & 1 & 1 & 3 & 3 & 1 & ${\frac {4\,\sqrt {21}}{21}} R_{{40}}$\\
1 & 1 & 1 & 5 & 5 & 1 & $ {\frac {20\,\sqrt {3927}}{1309}} R_{{40}} +{\frac {4\,\sqrt {51051}}{2431}} R_{{60}}$\\
1 & 1 & 1 & 5 & 5 & 2 & $ -{\frac {2\,\sqrt {2805}}{561}} R_{{40}} +{\frac {24\,\sqrt {36465}}{2431}} R_{{60}}$\\
3 & 3 & 1 & 3 & 3 & 1 & $  R_{{00}} +{\frac {6}{11}} R_{{40}} +{\frac {100\,\sqrt {13}}{429}} R_{{60}}$\\
3 & 3 & 1 & 5 & 5 & 1 & $ {\frac {60\,\sqrt {187}}{2431}} R_{{40}} +{\frac {42\,\sqrt {2431}}{2431}} R_{{60}} +{\frac {112\,\sqrt {11}}{429}} R_{{80}}$\\
3 & 3 & 1 & 5 & 5 & 2 & $ {\frac {12\,\sqrt {6545}}{1309}} R_{{40}} -{\frac {28\,\sqrt {85085}}{7293}} R_{{60}}$\\
5 & 5 & 1 & 5 & 5 & 1 & $  R_{{00}} +{\frac {132}{221}} R_{{40}} +{\frac {880\,\sqrt {13}}{3757}} R_{{60}} +{\frac {280\,\sqrt {17}}{3757}} R_{{80}} +{\frac {336\,\sqrt {21}}{3757}} R_{{10,0}}$\\
5 & 5 & 1 & 5 & 5 & 2 & $ -{\frac {24\,\sqrt {35}}{1547}} R_{{40}} -{\frac {120\,\sqrt {455}}{3757}} R_{{60}} +{\frac {2800\,\sqrt {595}}{214149}} R_{{80}}$\\
 &&&&&& $\qquad  +{\frac {88704\,\sqrt {15}}{356915}} R_{{10,0}}$\\
5 & 5 & 2 & 5 & 5 & 2 & $  R_{{00}} -{\frac {132}{221}} R_{{40}} +{\frac {352\,\sqrt {13}}{11271}} R_{{60}} +{\frac {7056\,\sqrt {17}}{71383}} R_{{80}} $\\
 &&&&&& $\qquad -{\frac {12096\,\sqrt {21}}{71383}} R_{{10,0}}$\\
\hline
\multicolumn{7}{c}{}\\[-5pt]
\multicolumn{7}{c}{$\Lambda_B=T_{2g}$}\\ \hline
2 & 2 & 1 & 2 & 2 & 1 & $  R_{{00}} -\frac{4}{7} R_{{40}}$\\
2 & 2 & 1 & 4 & 4 & 1 & $ -{\frac {20\,\sqrt {3}}{77}} R_{{40}} +{\frac {40\,\sqrt {39}}{143}} R_{{60}}$\\
2 & 2 & 1 & 6 & 6 & 1 & $ {\frac {20\,\sqrt {715}}{1001}} R_{{40}} -{\frac {12\,\sqrt {55}}{55}} R_{{60}} -{\frac {32\,\sqrt {12155}}{36465}} R_{{80}}$\\
2 & 2 & 1 & 6 & 6 & 2 & $ {\frac {190\,\sqrt {13}}{1001}} R_{{40}} +{\frac {8}{11}} R_{{60}} -{\frac {32\,\sqrt {221}}{663}} R_{{80}}$\\
4 & 4 & 1 & 4 & 4 & 1 & $  R_{{00}} -{\frac {54}{77}} R_{{40}} +{\frac {20\,\sqrt {13}}{143}} R_{{60}}$\\
4 & 4 & 1 & 6 & 6 & 1 & $ {\frac {4\,\sqrt {2145}}{1001}} R_{{40}} -{\frac {2\,\sqrt {165}}{187}} R_{{60}} -{\frac {144\,\sqrt {36465}}{46189}} R_{{80}} +{\frac {384\,\sqrt {5005}}{20995}} R_{{10,0}}$\\
4 & 4 & 1 & 6 & 6 & 2 & $ -{\frac {60\,\sqrt {39}}{1001}} R_{{40}} -{\frac {124\,\sqrt {3}}{187}} R_{{60}} +{\frac {64\,\sqrt {663}}{4199}} R_{{80}} +{\frac {192\,\sqrt {91}}{4199}} R_{{10,0}}$\\
6 & 6 & 1 & 6 & 6 & 1 & $  R_{{00}} -{\frac {32}{119}} R_{{40}} +{\frac {80\,\sqrt {13}}{323}} R_{{60}} -{\frac {920\,\sqrt {17}}{6783}} R_{{80}} -{\frac {720\,\sqrt {21}}{52003}} R_{{10,0}}$\\
 &&&&&& $\qquad  +{\frac {91608}{260015}} R_{{12,0}} -{\frac {5808\,\sqrt {1001}}{260015}} R_{{12,4}}$\\
6 & 6 & 1 & 6 & 6 & 2 & $ {\frac {40\,\sqrt {55}}{1309}} R_{{40}} +{\frac {120\,\sqrt {715}}{3553}} R_{{60}} +{\frac {80\,\sqrt {935}}{24871}} R_{{80}} -{\frac {4608\,\sqrt {1155}}{260015}} R_{{10,0}} $\\
 &&&&&& $\qquad -{\frac {13728\,\sqrt {55}}{260015}} R_{{12,0}} +{\frac {6336\,\sqrt {455}}{260015}} R_{{12,4}}$\\
6 & 6 & 2 & 6 & 6 & 2 & $  R_{{00}} +{\frac {632}{1309}} R_{{40}} -{\frac {480\,\sqrt {13}}{3553}} R_{{60}} +{\frac {80\,\sqrt {17}}{6783}} R_{{80}} +{\frac {1728\,\sqrt {21}}{52003}} R_{{10,0}}$\\
 &&&&&& $\qquad  -{\frac {29040}{52003}} R_{{12,0}} -{\frac {1056\,\sqrt {1001}}{52003}} R_{{12,4}}$\\
\hline
\multicolumn{7}{c}{}\\[-5pt]
\multicolumn{7}{c}{$\Lambda_B=T_{2u}$}\\ \hline
3 & 3 & 1 & 3 & 3 & 1 & $  R_{{00}} -\frac{2}{11} R_{{40}} -{\frac {60\,\sqrt {13}}{143}} R_{{60}}$\\
3 & 3 & 1 & 5 & 5 & 1 & $ -{\frac {20\,\sqrt {11}}{143}} R_{{40}} -{\frac {14\,\sqrt {143}}{143}} R_{{60}} +{\frac {112\,\sqrt {187}}{2431}} R_{{80}}$\\
5 & 5 & 1 & 5 & 5 & 1 & $  R_{{00}} +{\frac {4}{13}} R_{{40}} -{\frac {80\,\sqrt {13}}{221}} R_{{60}} -{\frac {280\,\sqrt {17}}{4199}} R_{{80}} -{\frac {432\,\sqrt {21}}{4199}} R_{{10,0}}$\\
\hline
\end{tabular}
\end{center}
\end{table}

\begin{table}[p]
\caption{
Box matrix elements $B^{(\Pvec\Lambda_B S a)}_{J'L'n';\ JLn}(E)$
for various irreps with $\Pvec=0$ and total spin $S=\frac{1}{2}$.
These quantities depend on $a$ only through $\svec_a$ and $u_a$.  
$R_{lm}$ is short hand
for $(\gamma \pi^{3/2}u_a^{l+1})^{-1}{\rm Re}\ {\cal Z}_{lm}(\svec_a,\gamma,u_a^2)$.
The Hermiticity of $B$ can be
used to obtain other elements that are not shown.
\label{tab:boxP02S1A}}
\begin{center}
\begin{tabular}{|ccc|ccc|l|}\hline
$J'$ & $L'$ & $n'$ & $J$ & $L$& $n$ & $\qquad\qquad u_a^{-(L'+L+1)}\ B$ \\ \hline
\multicolumn{7}{c}{}\\[-5pt]
\multicolumn{7}{c}{$\Lambda_B=G_{1g}$}\\ \hline
$\frac{1}{2}$ & 0 & 1 & $\frac{1}{2}$ & 0 & 1 & $ R_{00}$\\
$\frac{1}{2}$ & 0 & 1 & $\frac{7}{2}$ & 4 & 1 & $-\frac{4\,\sqrt {21}}{21} R_{40}$\\
$\frac{1}{2}$ & 0 & 1 & $\frac{9}{2}$ & 4 & 1 & $\frac{2\,\sqrt {105}}{21} R_{40}$\\
$\frac{1}{2}$ & 0 & 1 & $\frac{11}{2}$ & 6 & 1 & $\frac{4\,\sqrt {39}}{13} R_{60}$\\
$\frac{1}{2}$ & 0 & 1 & $\frac{13}{2}$ & 6 & 1 & $-\frac{2\,\sqrt {182}}{13} R_{60}$\\
$\frac{7}{2}$ & 4 & 1 & $\frac{7}{2}$ & 4 & 1 & $  R_{00} +\frac{6}{11} R_{40} +\frac{100\,\sqrt {13}}{429} R_{60}$\\
$\frac{7}{2}$ & 4 & 1 & $\frac{9}{2}$ & 4 & 1 & $ -\frac{12\,\sqrt {5}}{143} R_{40} -\frac{56\,\sqrt {65}}{429} R_{60} -\frac{224\,\sqrt {85}}{2431} R_{80}$\\
$\frac{7}{2}$ & 4 & 1 & $\frac{11}{2}$ & 6 & 1 & $ -\frac{300\,\sqrt {7}}{1001} R_{40} +\frac{14\,\sqrt {91}}{143} R_{60} -\frac{112\,\sqrt {119}}{7293} R_{80}$\\
$\frac{7}{2}$ & 4 & 1 & $\frac{13}{2}$ & 6 & 1 & $ \frac{20\,\sqrt {6}}{429} R_{40} -\frac{126\,\sqrt {78}}{2431} R_{60} +\frac{112\,\sqrt {102}}{4199} R_{80} +\frac{96\,\sqrt {14}}{323} R_{10,0}$\\
$\frac{9}{2}$ & 4 & 1 & $\frac{9}{2}$ & 4 & 1 & $  R_{00} +\frac{84}{143} R_{40} +\frac{128\,\sqrt {13}}{429} R_{60} +\frac{112\,\sqrt {17}}{2431} R_{80}$\\
$\frac{9}{2}$ & 4 & 1 & $\frac{11}{2}$ & 6 & 1 & $ \frac{24\,\sqrt {35}}{1001} R_{40} -\frac{56\,\sqrt {455}}{2431} R_{60} +\frac{1568\,\sqrt {595}}{138567} R_{80} +\frac{6048\,\sqrt {15}}{20995} R_{10,0}$\\
$\frac{9}{2}$ & 4 & 1 & $\frac{13}{2}$ & 6 & 1 & $ -\frac{64\,\sqrt {30}}{429} R_{40} +\frac{126\,\sqrt {390}}{2431} R_{60} -\frac{448\,\sqrt {510}}{46189} R_{80} -\frac{528\,\sqrt {70}}{20995} R_{10,0}$\\
$\frac{11}{2}$ & 6 & 1 & $\frac{11}{2}$ & 6 & 1 & $  R_{00} -\frac{84}{143} R_{40} -\frac{80\,\sqrt {13}}{2431} R_{60} +\frac{5880\,\sqrt {17}}{46189} R_{80}$\\
 &&&&&& $\qquad  -\frac{336\,\sqrt {21}}{4199} R_{10,0}$\\
$\frac{11}{2}$ & 6 & 1 & $\frac{13}{2}$ & 6 & 1 & $ \frac{30\,\sqrt {42}}{2431} R_{40} +\frac{80\,\sqrt {546}}{46189} R_{60} -\frac{720\,\sqrt {714}}{46189} R_{80} +\frac{55440\,\sqrt {2}}{96577} R_{10,0}$\\
 &&&&&& $\qquad  -\frac{4356\,\sqrt {42}}{37145} R_{12,0} +\frac{1848\,\sqrt {858}}{37145} R_{12,4}$\\
$\frac{13}{2}$ & 6 & 1 & $\frac{13}{2}$ & 6 & 1 & $  R_{00} -\frac{1458}{2431} R_{40} -\frac{1600\,\sqrt {13}}{46189} R_{60} +\frac{600\,\sqrt {17}}{4199} R_{80}$\\
 &&&&&& $\qquad  -\frac{10368\,\sqrt {21}}{96577} R_{10,0} +\frac{4356}{37145} R_{12,0} -\frac{264\,\sqrt {1001}}{37145} R_{12,4}$\\
\hline
\multicolumn{7}{c}{}\\[-5pt]
\multicolumn{7}{c}{$\Lambda_B=G_{1u}$}\\ \hline
$\frac{1}{2}$ & 1 & 1 & $\frac{1}{2}$ & 1 & 1 & $ R_{00}$\\
$\frac{1}{2}$ & 1 & 1 & $\frac{7}{2}$ & 3 & 1 & $-\frac{4\,\sqrt {21}}{21} R_{40}$\\
$\frac{1}{2}$ & 1 & 1 & $\frac{9}{2}$ & 5 & 1 & $\frac{2\,\sqrt {105}}{21} R_{40}$\\
$\frac{1}{2}$ & 1 & 1 & $\frac{11}{2}$ & 5 & 1 & $\frac{4\,\sqrt {39}}{13} R_{60}$\\
$\frac{7}{2}$ & 3 & 1 & $\frac{7}{2}$ & 3 & 1 & $  R_{00} +\frac{6}{11} R_{40} +\frac{100\,\sqrt {13}}{429} R_{60}$\\
$\frac{7}{2}$ & 3 & 1 & $\frac{9}{2}$ & 5 & 1 & $ -\frac{12\,\sqrt {5}}{143} R_{40} -\frac{56\,\sqrt {65}}{429} R_{60} -\frac{224\,\sqrt {85}}{2431} R_{80}$\\
$\frac{7}{2}$ & 3 & 1 & $\frac{11}{2}$ & 5 & 1 & $ -\frac{300\,\sqrt {7}}{1001} R_{40} +\frac{14\,\sqrt {91}}{143} R_{60} -\frac{112\,\sqrt {119}}{7293} R_{80}$\\
$\frac{9}{2}$ & 5 & 1 & $\frac{9}{2}$ & 5 & 1 & $  R_{00} +\frac{84}{143} R_{40} +\frac{128\,\sqrt {13}}{429} R_{60} +\frac{112\,\sqrt {17}}{2431} R_{80}$\\
$\frac{9}{2}$ & 5 & 1 & $\frac{11}{2}$ & 5 & 1 & $ \frac{24\,\sqrt {35}}{1001} R_{40} -\frac{56\,\sqrt {455}}{2431} R_{60} +\frac{1568\,\sqrt {595}}{138567} R_{80} +\frac{6048\,\sqrt {15}}{20995} R_{10,0}$\\
$\frac{11}{2}$ & 5 & 1 & $\frac{11}{2}$ & 5 & 1 & $  R_{00} -\frac{84}{143} R_{40} -\frac{80\,\sqrt {13}}{2431} R_{60} +\frac{5880\,\sqrt {17}}{46189} R_{80}$\\
 &&&&&& $\qquad  -\frac{336\,\sqrt {21}}{4199} R_{10,0}$\\
\hline
\end{tabular}
\end{center}
\end{table}

\begin{table}[p]
\caption{
Box matrix elements $B^{(\Pvec\Lambda_B S a)}_{J'L'n';\ JLn}(E)$
for various irreps with $\Pvec=0$ and total spin $S=1$.
These quantities depend on $a$ only through $\svec_a$ and $u_a$.  
$R_{lm}$ is short hand
for $(\gamma \pi^{3/2}u_a^{l+1})^{-1}{\rm Re}\ {\cal Z}_{lm}(\svec_a,\gamma,u_a^2)$.
The Hermiticity of $B$ can be
used to obtain other elements that are not shown.
\label{tab:boxP0S1A}}
\begin{center}
\begin{tabular}{|ccc|ccc|l|}\hline
$J'$ & $L'$ & $n'$ & $J$ & $L$& $n$ & $\qquad\qquad u_a^{-(L'+L+1)}\ B$ \\ \hline
\multicolumn{7}{c}{}\\[-5pt]
\multicolumn{7}{c}{$\Lambda_B=A_{1g}$}\\ \hline
$4$ & 4 & 1 & $4$ & 4 & 1 & $  R_{00} +\frac{54}{143} R_{40} -\frac{4\,\sqrt {13}}{143} R_{60} -\frac{448\,\sqrt {17}}{2431} R_{80}$\\
$4$ & 4 & 1 & $6$ & 6 & 1 & $ -\frac{12\,\sqrt {65}}{143} R_{40} +\frac{42\,\sqrt {5}}{187} R_{60} +\frac{112\,\sqrt {1105}}{46189} R_{80} +\frac{576\,\sqrt {1365}}{20995} R_{10,0}$\\
$6$ & 6 & 1 & $6$ & 6 & 1 & $  R_{00} -\frac{96}{187} R_{40} -\frac{80\,\sqrt {13}}{3553} R_{60} +\frac{120\,\sqrt {17}}{3553} R_{80}$\\
 &&&&&& $\qquad  +\frac{624\,\sqrt {21}}{7429} R_{10,0} -\frac{26136}{37145} R_{12,0} +\frac{1584\,\sqrt {1001}}{37145} R_{12,4}$\\
\hline
\multicolumn{7}{c}{}\\[-5pt]
\multicolumn{7}{c}{$\Lambda_B=A_{1u}$}\\ \hline
$0$ & 1 & 1 & $0$ & 1 & 1 & $ R_{00}$\\
$0$ & 1 & 1 & $4$ & 3 & 1 & $-\frac{4\,\sqrt {21}}{21} R_{40}$\\
$0$ & 1 & 1 & $4$ & 5 & 1 & $\frac{2\,\sqrt {105}}{21} R_{40}$\\
$0$ & 1 & 1 & $6$ & 5 & 1 & $\frac{4\,\sqrt {39}}{13} R_{60}$\\
$4$ & 3 & 1 & $4$ & 3 & 1 & $  R_{00} +\frac{6}{11} R_{40} +\frac{100\,\sqrt {13}}{429} R_{60}$\\
$4$ & 3 & 1 & $4$ & 5 & 1 & $ -\frac{12\,\sqrt {5}}{143} R_{40} -\frac{56\,\sqrt {65}}{429} R_{60} -\frac{224\,\sqrt {85}}{2431} R_{80}$\\
$4$ & 3 & 1 & $6$ & 5 & 1 & $ -\frac{300\,\sqrt {7}}{1001} R_{40} +\frac{14\,\sqrt {91}}{143} R_{60} -\frac{112\,\sqrt {119}}{7293} R_{80}$\\
$4$ & 5 & 1 & $4$ & 5 & 1 & $  R_{00} +\frac{84}{143} R_{40} +\frac{128\,\sqrt {13}}{429} R_{60} +\frac{112\,\sqrt {17}}{2431} R_{80}$\\
$4$ & 5 & 1 & $6$ & 5 & 1 & $ \frac{24\,\sqrt {35}}{1001} R_{40} -\frac{56\,\sqrt {455}}{2431} R_{60} +\frac{1568\,\sqrt {595}}{138567} R_{80} +\frac{6048\,\sqrt {15}}{20995} R_{10,0}$\\
$6$ & 5 & 1 & $6$ & 5 & 1 & $  R_{00} -\frac{84}{143} R_{40} -\frac{80\,\sqrt {13}}{2431} R_{60} +\frac{5880\,\sqrt {17}}{46189} R_{80}$\\
 &&&&&& $\qquad  -\frac{336\,\sqrt {21}}{4199} R_{10,0}$\\
\hline
\multicolumn{7}{c}{}\\[-5pt]
\multicolumn{7}{c}{$\Lambda_B=A_{2g}$}\\ \hline
$3$ & 2 & 1 & $3$ & 2 & 1 & $  R_{00} -\frac{4}{7} R_{40}$\\
$3$ & 2 & 1 & $3$ & 4 & 1 & $ \frac{20\,\sqrt {3}}{77} R_{40} -\frac{40\,\sqrt {39}}{143} R_{60}$\\
$3$ & 2 & 1 & $6$ & 6 & 1 & $ \frac{40\,\sqrt {143}}{1001} R_{40} +\frac{4\,\sqrt {11}}{11} R_{60} -\frac{32\,\sqrt {2431}}{2431} R_{80}$\\
$3$ & 2 & 1 & $7$ & 6 & 1 & $ \frac{90\,\sqrt {65}}{1001} R_{40} -\frac{32\,\sqrt {5}}{55} R_{60} -\frac{32\,\sqrt {1105}}{3315} R_{80}$\\
$3$ & 4 & 1 & $3$ & 4 & 1 & $  R_{00} -\frac{54}{77} R_{40} +\frac{20\,\sqrt {13}}{143} R_{60}$\\
$3$ & 4 & 1 & $6$ & 6 & 1 & $ \frac{20\,\sqrt {429}}{1001} R_{40} +\frac{2\,\sqrt {33}}{11} R_{60} -\frac{16\,\sqrt {7293}}{2431} R_{80}$\\
$3$ & 4 & 1 & $7$ & 6 & 1 & $ -\frac{4\,\sqrt {195}}{1001} R_{40} +\frac{24\,\sqrt {15}}{187} R_{60} +\frac{32\,\sqrt {3315}}{4199} R_{80} -\frac{1344\,\sqrt {455}}{20995} R_{10,0}$\\
$6$ & 6 & 1 & $6$ & 6 & 1 & $  R_{00} +\frac{32}{119} R_{40} -\frac{80\,\sqrt {13}}{323} R_{60} -\frac{40\,\sqrt {17}}{2261} R_{80}$\\
 &&&&&& $\qquad  +\frac{5616\,\sqrt {21}}{52003} R_{10,0} -\frac{11880}{52003} R_{12,0} -\frac{1584\,\sqrt {1001}}{52003} R_{12,4}$\\
$6$ & 6 & 1 & $7$ & 6 & 1 & $ \frac{72\,\sqrt {55}}{1309} R_{40} +\frac{40\,\sqrt {715}}{3553} R_{60} +\frac{640\,\sqrt {935}}{74613} R_{80} -\frac{3168\,\sqrt {1155}}{260015} R_{10,0}$\\
 &&&&&& $\qquad  -\frac{20592\,\sqrt {55}}{260015} R_{12,0} +\frac{1056\,\sqrt {455}}{52003} R_{12,4}$\\
$7$ & 6 & 1 & $7$ & 6 & 1 & $  R_{00} -\frac{72}{1309} R_{40} +\frac{1280\,\sqrt {13}}{3553} R_{60} -\frac{240\,\sqrt {17}}{2261} R_{80}$\\
 &&&&&& $\qquad  -\frac{4608\,\sqrt {21}}{52003} R_{10,0} +\frac{5808}{260015} R_{12,0} -\frac{3168\,\sqrt {1001}}{260015} R_{12,4}$\\
\hline
\end{tabular}
\end{center}
\end{table}

\begin{table}[p]
\caption{
Selected box matrix elements $B^{(\Pvec\Lambda_B S a)}_{J'L'n';\ JLn}(E)$
for $\Pvec=(2\pi/L)(0,0,n)$ and total spin $S=0$.
These quantities depend on $a$ only through $\svec_a$ and $u_a$.  
$R_{lm}$ is short hand
for $(\gamma \pi^{3/2}u_a^{l+1})^{-1}{\rm Re}\ {\cal Z}_{lm}(\svec_a,\gamma,u_a^2)$.
The Hermiticity of $B$ can be
used to obtain other elements that are not shown.
\label{tab:boxP1S0A}}
\begin{center}
\begin{tabular}{|ccc|ccc|l|}\hline
$J'$ & $L'$ & $n'$ & $J$ & $L$& $n$ & $\qquad\qquad u_a^{-(L'+L+1)}\ B$ \\ \hline
\multicolumn{7}{c}{}\\[-5pt]
\multicolumn{7}{c}{$\Lambda_B=B_{1}$\qquad\mbox{(partial)}}\\ \hline
$2$ & 2 & 1 & $2$ & 2 & 1 & $  R_{00} -\frac{2\,\sqrt {5}}{7} R_{20} +\frac{1}{7} R_{40} +\frac{\sqrt {70}}{7} R_{44}$\\
$2$ & 2 & 1 & $3$ & 3 & 1 & $ \frac{\sqrt {21}}{7} R_{10} -\frac{2}{3} R_{30} +\frac{5\,\sqrt {77}}{231} R_{50} +\frac{\sqrt {110}}{11} R_{54}$\\
$2$ & 2 & 1 & $4$ & 4 & 1 & $ \frac{\sqrt {15}}{7} R_{20} -\frac{30\,\sqrt {3}}{77} R_{40} -\frac{2\,\sqrt {210}}{77} R_{44} +\frac{5\,\sqrt {39}}{143} R_{60}$\\
 &&&&&& $\qquad  +\frac{5\,\sqrt {546}}{143} R_{64}$\\
$2$ & 2 & 1 & $5$ & 5 & 1 & $ \frac{5\,\sqrt {11}}{33} R_{30} -\frac{10\,\sqrt {7}}{39} R_{50} -\frac{2\,\sqrt {10}}{13} R_{54} +\frac{\sqrt {1155}}{143} R_{70}$\\
 &&&&&& $\qquad  +\frac{3\,\sqrt {10}}{13} R_{74}$\\
$2$ & 2 & 1 & $6$ & 6 & 1 & $ \frac{15\,\sqrt {143}}{143} R_{44} -\frac{2\,\sqrt {55}}{55} R_{64} +\frac{\sqrt {1105}}{1105} R_{84} +\frac{2\,\sqrt {119}}{17} R_{88}$\\
$2$ & 2 & 1 & $6$ & 6 & 2 & $ \frac{5\,\sqrt {182}}{143} R_{40} +\frac{\sqrt {65}}{143} R_{44} -\frac{2\,\sqrt {14}}{11} R_{60} -\frac{6}{11} R_{64}$\\
 &&&&&& $\qquad  +\frac{\sqrt {3094}}{221} R_{80} +\frac{3\,\sqrt {2431}}{221} R_{84}$\\
$3$ & 3 & 1 & $3$ & 3 & 1 & $  R_{00} -\frac{7}{11} R_{40} +\frac{\sqrt {70}}{11} R_{44} +\frac{10\,\sqrt {13}}{143} R_{60}$\\
 &&&&&& $\qquad  +\frac{10\,\sqrt {182}}{143} R_{64}$\\
$3$ & 3 & 1 & $4$ & 4 & 1 & $ \frac{2\,\sqrt {7}}{7} R_{10} -\frac{\sqrt {3}}{11} R_{30} -\frac{40\,\sqrt {231}}{1001} R_{50} +\frac{4\,\sqrt {330}}{143} R_{54}$\\
 &&&&&& $\qquad  +\frac{7\,\sqrt {35}}{143} R_{70} +\frac{7\,\sqrt {330}}{143} R_{74}$\\
$3$ & 3 & 1 & $5$ & 5 & 1 & $ \frac{\sqrt {55}}{11} R_{20} -\frac{10\,\sqrt {11}}{143} R_{40} -\frac{2\,\sqrt {770}}{143} R_{44} -\frac{7\,\sqrt {143}}{143} R_{60}$\\
 &&&&&& $\qquad  +\frac{\sqrt {2002}}{143} R_{64} +\frac{56\,\sqrt {187}}{2431} R_{80} +\frac{12\,\sqrt {238}}{221} R_{84}$\\
$3$ & 3 & 1 & $6$ & 6 & 1 & $ \frac{\sqrt {91}}{13} R_{54} -\frac{6\,\sqrt {91}}{221} R_{74} +\frac{\sqrt {133}}{323} R_{94} +\frac{14\,\sqrt {323}}{323} R_{98}$\\
$3$ & 3 & 1 & $6$ & 6 & 2 & $ \frac{20\,\sqrt {182}}{429} R_{30} -\frac{7\,\sqrt {286}}{429} R_{50} -\frac{\sqrt {5005}}{143} R_{54} -\frac{70\,\sqrt {390}}{2431} R_{70}$\\
 &&&&&& $\qquad  +\frac{6\,\sqrt {5005}}{2431} R_{74} +\frac{63\,\sqrt {494}}{4199} R_{90} +\frac{3\,\sqrt {7315}}{323} R_{94}$\\
$4$ & 4 & 1 & $4$ & 4 & 1 & $  R_{00} +\frac{8\,\sqrt {5}}{77} R_{20} -\frac{27}{91} R_{40} +\frac{81\,\sqrt {70}}{1001} R_{44}$\\
 &&&&&& $\qquad  -\frac{2\,\sqrt {13}}{13} R_{60} +\frac{6\,\sqrt {182}}{143} R_{64} +\frac{196\,\sqrt {17}}{2431} R_{80} +\frac{42\,\sqrt {2618}}{2431} R_{84}$\\
$4$ & 4 & 1 & $5$ & 5 & 1 & $ \frac{\sqrt {77}}{11} R_{10} +\frac{2\,\sqrt {33}}{143} R_{30} -\frac{\sqrt {21}}{13} R_{50} +\frac{\sqrt {30}}{13} R_{54}$\\
 &&&&&& $\qquad  -\frac{64\,\sqrt {385}}{2431} R_{70} +\frac{20\,\sqrt {30}}{221} R_{74} +\frac{252\,\sqrt {4389}}{46189} R_{90} $\\
 &&&&&& $\qquad +\frac{42\,\sqrt {7410}}{4199} R_{94}$\\
$4$ & 4 & 1 & $6$ & 6 & 1 & $ -\frac{4\,\sqrt {429}}{143} R_{44} +\frac{9\,\sqrt {165}}{187} R_{64} -\frac{18\,\sqrt {3315}}{4199} R_{84} $\\
 &&&&&& $\qquad  -\frac{12\,\sqrt {357}}{323} R_{88}$\\
 &&&&&& $\qquad  +\frac{7\,\sqrt {3}}{323} R_{10,4} +\frac{42\,\sqrt {17}}{323} R_{10,8}$\\
$4$ & 4 & 1 & $6$ & 6 & 2 & $ \frac{2\,\sqrt {2730}}{143} R_{20} -\frac{4\,\sqrt {195}}{143} R_{44} -\frac{\sqrt {42}}{17} R_{60} +\frac{23\,\sqrt {3}}{187} R_{64}$\\
 &&&&&& $\qquad  -\frac{18\,\sqrt {9282}}{3553} R_{80} +\frac{222\,\sqrt {7293}}{46189} R_{84} +\frac{315\,\sqrt {26}}{4199} R_{10,0} $\\
 &&&&&& $\qquad  +\frac{21\,\sqrt {165}}{323} R_{10,4}$\\
$5$ & 5 & 1 & $5$ & 5 & 1 & $  R_{00} +\frac{2\,\sqrt {5}}{13} R_{20} -\frac{1}{13} R_{40} +\frac{\sqrt {70}}{13} R_{44}$\\
 &&&&&& $\qquad  -\frac{24\,\sqrt {13}}{221} R_{60} +\frac{8\,\sqrt {182}}{221} R_{64} -\frac{28\,\sqrt {17}}{247} R_{80} +\frac{42\,\sqrt {2618}}{4199} R_{84}$\\
 &&&&&& $\qquad  +\frac{360\,\sqrt {21}}{4199} R_{10,0} +\frac{36\,\sqrt {10010}}{4199} R_{10,4}$\\
\hline
\end{tabular}
\end{center}
\end{table}

\begin{table}[p]
\caption{
Selected box matrix elements $B^{(\Pvec\Lambda_B S a)}_{J'L'n';\ JLn}(E)$
for $\Pvec=(2\pi/L)(0,0,n)$ and total spin $S=\frac{1}{2}$.
These quantities depend on $a$ only through $\svec_a$ and $u_a$.  
$R_{lm}$ is short hand
for $(\gamma \pi^{3/2}u_a^{l+1})^{-1}{\rm Re}\ {\cal Z}_{lm}(\svec_a,\gamma,u_a^2)$.
The Hermiticity of $B$ can be
used to obtain other elements that are not shown.
\label{tab:boxP1S0B}}
\begin{center}
\begin{tabular}{|ccc|ccc|l|}\hline
$J'$ & $L'$ & $n'$ & $J$ & $L$& $n$ & $\qquad u_a^{-(L'+L+1)}\ B$ \\ \hline
\multicolumn{7}{c}{}\\[-5pt]
\multicolumn{7}{c}{$\Lambda_B=G_{2}$\qquad\mbox{(partial)}}\\ \hline
$\frac{3}{2}$ & 1 & 1 & $\frac{3}{2}$ & 1 & 1 & $  R_{00} -\frac{\sqrt {5}}{5} R_{20}$\\
$\frac{3}{2}$ & 1 & 1 & $\frac{3}{2}$ & 2 & 1 & $ \frac{\sqrt {3}}{5} R_{10} -\frac{3\,\sqrt {7}}{35} R_{30}$\\
$\frac{3}{2}$ & 1 & 1 & $\frac{5}{2}$ & 2 & 2 & $ \frac{2\,\sqrt {3}}{5} R_{10} -\frac{6\,\sqrt {7}}{35} R_{30}$\\
$\frac{3}{2}$ & 1 & 1 & $\frac{5}{2}$ & 3 & 1 & $-\frac{2\,\sqrt {14}}{7} R_{44}$\\
$\frac{3}{2}$ & 1 & 1 & $\frac{5}{2}$ & 3 & 2 & $ \frac{6\,\sqrt {5}}{35} R_{20} -\frac{2}{7} R_{40}$\\
$\frac{3}{2}$ & 1 & 1 & $\frac{7}{2}$ & 3 & 1 & $-\frac{2\,\sqrt {21}}{21} R_{44}$\\
$\frac{3}{2}$ & 1 & 1 & $\frac{7}{2}$ & 3 & 2 & $ \frac{3\,\sqrt {2}}{7} R_{20} -\frac{\sqrt {10}}{7} R_{40}$\\
$\frac{3}{2}$ & 1 & 1 & $\frac{7}{2}$ & 4 & 1 & $-\frac{2\,\sqrt {231}}{33} R_{54}$\\
$\frac{3}{2}$ & 1 & 1 & $\frac{7}{2}$ & 4 & 2 & $ \frac{\sqrt {70}}{21} R_{30} -\frac{\sqrt {110}}{33} R_{50}$\\
$\frac{3}{2}$ & 1 & 1 & $\frac{9}{2}$ & 4 & 1 & $-\frac{2\,\sqrt {66}}{33} R_{54}$\\
$\frac{3}{2}$ & 1 & 1 & $\frac{9}{2}$ & 4 & 2 & $ \frac{2\,\sqrt {35}}{21} R_{30} -\frac{2\,\sqrt {55}}{33} R_{50}$\\
$\frac{3}{2}$ & 1 & 1 & $\frac{9}{2}$ & 5 & 1 & $ \frac{2\,\sqrt {6}}{33} R_{44} -\frac{6\,\sqrt {390}}{143} R_{64}$\\
$\frac{3}{2}$ & 1 & 1 & $\frac{9}{2}$ & 5 & 2 & $ \frac{2\,\sqrt {5}}{11} R_{40} -\frac{6\,\sqrt {65}}{143} R_{60}$\\
$\frac{3}{2}$ & 1 & 1 & $\frac{11}{2}$ & 5 & 1 & $ \frac{1}{11} R_{44} -\frac{9\,\sqrt {65}}{143} R_{64}$\\
$\frac{3}{2}$ & 1 & 1 & $\frac{11}{2}$ & 5 & 2 & $ \frac{\sqrt {35}}{11} R_{40} -\frac{3\,\sqrt {455}}{143} R_{60}$\\
$\frac{3}{2}$ & 1 & 1 & $\frac{11}{2}$ & 5 & 3 & $ \frac{\sqrt {165}}{11} R_{44} -\frac{\sqrt {429}}{143} R_{64}$\\
$\frac{3}{2}$ & 1 & 1 & $\frac{11}{2}$ & 6 & 1 & $ \frac{9\,\sqrt {11}}{143} R_{54} -\frac{3\,\sqrt {11}}{13} R_{74}$\\
$\frac{3}{2}$ & 1 & 1 & $\frac{11}{2}$ & 6 & 2 & $ \frac{3\,\sqrt {385}}{143} R_{50} -\frac{\sqrt {21}}{13} R_{70}$\\
$\frac{3}{2}$ & 1 & 1 & $\frac{11}{2}$ & 6 & 3 & $ \frac{\sqrt {15}}{13} R_{54} -\frac{\sqrt {15}}{65} R_{74}$\\
$\frac{3}{2}$ & 1 & 1 & $\frac{13}{2}$ & 6 & 2 & $ \frac{6\,\sqrt {11}}{143} R_{54} -\frac{2\,\sqrt {11}}{13} R_{74}$\\
$\frac{3}{2}$ & 1 & 1 & $\frac{13}{2}$ & 6 & 3 & $ \frac{6\,\sqrt {154}}{143} R_{50} -\frac{2\,\sqrt {210}}{65} R_{70}$\\
$\frac{3}{2}$ & 1 & 1 & $\frac{13}{2}$ & 6 & 4 & $ \frac{6\,\sqrt {5}}{13} R_{54} -\frac{6\,\sqrt {5}}{65} R_{74}$\\
$\frac{3}{2}$ & 2 & 1 & $\frac{3}{2}$ & 2 & 1 & $  R_{00} -\frac{\sqrt {5}}{5} R_{20}$\\
$\frac{3}{2}$ & 2 & 1 & $\frac{5}{2}$ & 2 & 1 & $-\frac{2\,\sqrt {14}}{7} R_{44}$\\
$\frac{3}{2}$ & 2 & 1 & $\frac{5}{2}$ & 2 & 2 & $ \frac{6\,\sqrt {5}}{35} R_{20} -\frac{2}{7} R_{40}$\\
$\frac{3}{2}$ & 2 & 1 & $\frac{5}{2}$ & 3 & 2 & $ \frac{2\,\sqrt {3}}{5} R_{10} -\frac{6\,\sqrt {7}}{35} R_{30}$\\
$\frac{3}{2}$ & 2 & 1 & $\frac{7}{2}$ & 3 & 1 & $-\frac{2\,\sqrt {231}}{33} R_{54}$\\
$\frac{3}{2}$ & 2 & 1 & $\frac{7}{2}$ & 3 & 2 & $ \frac{\sqrt {70}}{21} R_{30} -\frac{\sqrt {110}}{33} R_{50}$\\
$\frac{3}{2}$ & 2 & 1 & $\frac{7}{2}$ & 4 & 1 & $-\frac{2\,\sqrt {21}}{21} R_{44}$\\
$\frac{3}{2}$ & 2 & 1 & $\frac{7}{2}$ & 4 & 2 & $ \frac{3\,\sqrt {2}}{7} R_{20} -\frac{\sqrt {10}}{7} R_{40}$\\
$\frac{3}{2}$ & 2 & 1 & $\frac{9}{2}$ & 4 & 1 & $ \frac{2\,\sqrt {6}}{33} R_{44} -\frac{6\,\sqrt {390}}{143} R_{64}$\\
$\frac{3}{2}$ & 2 & 1 & $\frac{9}{2}$ & 4 & 2 & $ \frac{2\,\sqrt {5}}{11} R_{40} -\frac{6\,\sqrt {65}}{143} R_{60}$\\
$\frac{3}{2}$ & 2 & 1 & $\frac{9}{2}$ & 5 & 1 & $-\frac{2\,\sqrt {66}}{33} R_{54}$\\
\hline
\end{tabular}
\end{center}
\end{table}

\begin{table}[p]
\caption{
Selected box matrix elements $B^{(\Pvec\Lambda_B S a)}_{J'L'n';\ JLn}(E)$
for $\Pvec=(2\pi/L)(0,n,n)$ and total spin $S=0$.
These quantities depend on $a$ only through $\svec_a$ and $u_a$.  
$R_{lm}$ is short hand
for $(\gamma \pi^{3/2}u_a^{l+1})^{-1}{\rm Re}\ {\cal Z}_{lm}(\svec_a,\gamma,u_a^2)$, and
$I_{lm}$ is used to abbreviate $(\gamma \pi^{3/2}u_a^{l+1})^{-1}{\rm Im}\ {\cal Z}_{lm}(\svec_a,\gamma,u_a^2)$.
The Hermiticity of $B$ can be
used to obtain other elements that are not shown.
\label{tab:boxP2S0A}}
\begin{center}
\begin{tabular}{|ccc|ccc|l|}\hline
$J'$ & $L'$ & $n'$ & $J$ & $L$& $n$ & $\qquad u_a^{-(L'+L+1)}\ B$ \\ \hline
\multicolumn{7}{c}{}\\[-5pt]
\multicolumn{7}{c}{$\Lambda_B=A_{2}$\qquad\mbox{(partial)}}\\ \hline
$2$ & 2 & 1 & $2$ & 2 & 1 & $ \frac{\sqrt {30}}{7} I_{21} -\frac{8\,\sqrt {5}}{7} I_{41} + R_{00} -\frac{2\,\sqrt {5}}{7} R_{20}$\\
 &&&&&& $\qquad  -\frac{4}{7} R_{40} +\frac{2\,\sqrt {10}}{7} R_{42}$\\
$2$ & 2 & 1 & $3$ & 3 & 1 & $ -\frac{\sqrt {42}}{7}i R_{10} -\frac{\sqrt {2}}{6}i R_{30} -\frac{\sqrt {15}}{3}i R_{32} +\frac{25\,\sqrt {154}}{462}i R_{50}$\\
 &&&&&& $\qquad  -\frac{2\,\sqrt {165}}{33}i R_{52} -\frac{\sqrt {55}}{11}i R_{54}$\\
$2$ & 2 & 1 & $4$ & 4 & 1 & $ -\frac{2\,\sqrt {105}}{11} I_{41} +\frac{15\,\sqrt {26}}{143} I_{61} -\frac{18\,\sqrt {65}}{143} I_{63} -\frac{\sqrt {105}}{7} R_{20}$\\
 &&&&&& $\qquad  -\frac{2\,\sqrt {210}}{77} R_{42} -\frac{32\,\sqrt {65}}{143} R_{62}$\\
$2$ & 2 & 1 & $4$ & 4 & 2 & $ -\frac{2\,\sqrt {10}}{7}i I_{21} +\frac{26\,\sqrt {15}}{77}i I_{41} +\frac{5\,\sqrt {182}}{143}i I_{61} -\frac{6\,\sqrt {455}}{143}i I_{63}$\\
 &&&&&& $\qquad  -\frac{\sqrt {15}}{7}i R_{20} +\frac{20\,\sqrt {3}}{77}i R_{40} +\frac{4\,\sqrt {30}}{77}i R_{42} -\frac{40\,\sqrt {39}}{143}i R_{60}$\\
 &&&&&& $\qquad  -\frac{16\,\sqrt {455}}{143}i R_{62}$\\
$2$ & 2 & 1 & $5$ & 5 & 1 & $ \frac{5\,\sqrt {22}}{33}i R_{30} -\frac{2\,\sqrt {165}}{33}i R_{32} +\frac{5\,\sqrt {14}}{156}i R_{50} +\frac{\sqrt {15}}{39}i R_{52}$\\
 &&&&&& $\qquad  -\frac{11\,\sqrt {5}}{26}i R_{54} -\frac{7\,\sqrt {2310}}{572}i R_{70} +\frac{51\,\sqrt {110}}{572}i R_{72} -\frac{3\,\sqrt {5}}{26}i R_{74}$\\
 &&&&&& $\qquad  -\frac{3\,\sqrt {130}}{52}i R_{76}$\\
$2$ & 2 & 1 & $5$ & 5 & 2 & $ -\frac{5\,\sqrt {66}}{66} R_{30} -\frac{\sqrt {55}}{11} R_{32} +\frac{5\,\sqrt {42}}{39} R_{50} +\frac{2\,\sqrt {5}}{13} R_{52}$\\
 &&&&&& $\qquad  -\frac{2\,\sqrt {15}}{13} R_{54} -\frac{3\,\sqrt {770}}{286} R_{70} -\frac{\sqrt {330}}{286} R_{72} +\frac{3\,\sqrt {15}}{13} R_{74}$\\
 &&&&&& $\qquad  +\frac{\sqrt {390}}{26} R_{76}$\\
$2$ & 2 & 1 & $6$ & 6 & 1 & $ \frac{2\,\sqrt {165}}{55} I_{61} -\frac{\sqrt {66}}{11} I_{63} +\frac{7\,\sqrt {85085}}{1105} I_{81} +\frac{\sqrt {663}}{13} I_{83}$\\
 &&&&&& $\qquad  +\frac{15\,\sqrt {10010}}{2002} R_{40} -\frac{30\,\sqrt {1001}}{1001} R_{42} +\frac{\sqrt {770}}{55} R_{60} +\frac{16\,\sqrt {66}}{165} R_{62}$\\
 &&&&&& $\qquad  -\frac{3\,\sqrt {170170}}{2431} R_{80} +\frac{32\,\sqrt {4862}}{2431} R_{82} -\frac{19\,\sqrt {1105}}{1105} R_{84}$\\
$2$ & 2 & 1 & $6$ & 6 & 2 & $ -\frac{20\,\sqrt {273}}{143} I_{41} +\frac{9\,\sqrt {10}}{55} I_{61} -\frac{2}{11} I_{63} -\frac{22\,\sqrt {46410}}{1105} I_{81}$\\
 &&&&&& $\qquad  -\frac{10\,\sqrt {4862}}{221} I_{83} +\frac{5\,\sqrt {546}}{143} R_{42} -\frac{8}{55} R_{62} -\frac{4\,\sqrt {23205}}{1105} R_{80}$\\
 &&&&&& $\qquad  +\frac{6\,\sqrt {72930}}{1105} R_{84}$\\
$2$ & 2 & 1 & $6$ & 6 & 3 & $ \frac{8\,\sqrt {910}}{143}i I_{41} +\frac{2\,\sqrt {3}}{11}i I_{61} +\frac{\sqrt {30}}{11}i I_{63} -\frac{3\,\sqrt {1547}}{221}i I_{81}$\\
 &&&&&& $\qquad  +\frac{\sqrt {36465}}{221}i I_{83} +\frac{5\,\sqrt {182}}{154}i R_{40} +\frac{2\,\sqrt {455}}{1001}i R_{42} -\frac{5\,\sqrt {14}}{11}i R_{60}$\\
 &&&&&& $\qquad  -\frac{16\,\sqrt {30}}{55}i R_{62} +\frac{\sqrt {3094}}{221}i R_{80} -\frac{3\,\sqrt {2431}}{221}i R_{84}$\\
$3$ & 3 & 1 & $3$ & 3 & 1 & $  R_{00} -\frac{2}{11} R_{40} +\frac{5\,\sqrt {10}}{11} R_{42} -\frac{60\,\sqrt {13}}{143} R_{60}$\\
 &&&&&& $\qquad  -\frac{8\,\sqrt {1365}}{429} R_{62}$\\
$3$ & 3 & 1 & $4$ & 4 & 1 & $ -\frac{\sqrt {2}}{2}i R_{10} -\frac{\sqrt {42}}{11}i R_{30} +\frac{5\,\sqrt {66}}{572}i R_{50} -\frac{6\,\sqrt {385}}{143}i R_{52}$\\
 &&&&&& $\qquad  +\frac{5\,\sqrt {1155}}{286}i R_{54} +\frac{119\,\sqrt {10}}{572}i R_{70} -\frac{3\,\sqrt {210}}{52}i R_{72}$\\
 &&&&&& $\qquad  -\frac{\sqrt {1155}}{286}i R_{74} +\frac{\sqrt {30030}}{572}i R_{76}$\\
$3$ & 3 & 1 & $4$ & 4 & 2 & $ -\frac{3\,\sqrt {14}}{14} R_{10} +\frac{2\,\sqrt {6}}{11} R_{30} -\frac{2\,\sqrt {5}}{11} R_{32} -\frac{5\,\sqrt {462}}{364} R_{50}$\\
 &&&&&& $\qquad  -\frac{20\,\sqrt {55}}{143} R_{52} +\frac{\sqrt {165}}{286} R_{54} +\frac{21\,\sqrt {70}}{572} R_{70} +\frac{7\,\sqrt {30}}{572} R_{72}$\\
 &&&&&& $\qquad  -\frac{21\,\sqrt {165}}{286} R_{74} -\frac{7\,\sqrt {4290}}{572} R_{76}$\\
\hline
\end{tabular}
\end{center}
\end{table}

\begin{table}[p]
\caption{
Selected box matrix elements $B^{(\Pvec\Lambda_B S a)}_{J'L'n';\ JLn}(E)$
for $\Pvec=(2\pi/L)(0,n,n)$ and total spin $S=\frac{1}{2}$.
These quantities depend on $a$ only through $\svec_a$ and $u_a$.  
$R_{lm}$ is short hand
for $(\gamma \pi^{3/2}u_a^{l+1})^{-1}{\rm Re}\ {\cal Z}_{lm}(\svec_a,\gamma,u_a^2)$, and
$I_{lm}$ is used to abbreviate $(\gamma \pi^{3/2}u_a^{l+1})^{-1}{\rm Im}\ {\cal Z}_{lm}(\svec_a,\gamma,u_a^2)$.
The Hermiticity of $B$ can be
used to obtain other elements that are not shown.
\label{tab:boxP22S1A}}
\begin{center}
\begin{tabular}{|ccc|ccc|l|}\hline
$J'$ & $L'$ & $n'$ & $J$ & $L$& $n$ & $\qquad u_a^{-(L'+L+1)}\ B$ \\ \hline
\multicolumn{7}{c}{}\\[-5pt]
\multicolumn{7}{c}{$\Lambda_B=G$\qquad\mbox{(partial)}}\\ \hline
$\frac{5}{2}$ & 2 & 2 & $\frac{9}{2}$ & 5 & 4 & $ -\frac{3\,\sqrt {105}}{308}i R_{30} -\frac{13\,\sqrt {14}}{924}i R_{32} -\frac{7\,\sqrt {165}}{286}i R_{50} +\frac{95\,\sqrt {154}}{3003}i R_{52}$\\
 &&&&&& $\qquad  -\frac{25\,\sqrt {462}}{2002}i R_{54} +\frac{915}{2288}i R_{70} +\frac{375\,\sqrt {21}}{16016}i R_{72}$\\
 &&&&&& $\qquad   -\frac{675\,\sqrt {462}}{16016}i R_{74}+\frac{15\,\sqrt {3003}}{2288}i R_{76}$\\
$\frac{5}{2}$ & 2 & 2 & $\frac{9}{2}$ & 5 & 5 & $ -\frac{23\,\sqrt {30}}{924} R_{30} -\frac{95}{462} R_{32} -\frac{2\,\sqrt {2310}}{3003} R_{50} +\frac{2\,\sqrt {11}}{429} R_{52}$\\
 &&&&&& $\qquad  +\frac{16\,\sqrt {33}}{429} R_{54} +\frac{135\,\sqrt {14}}{2288} R_{70} +\frac{435\,\sqrt {6}}{2288} R_{72}$\\
 &&&&&& $\qquad   +\frac{105\,\sqrt {33}}{1144} R_{74}+\frac{45\,\sqrt {858}}{2288} R_{76}$\\
$\frac{5}{2}$ & 2 & 2 & $\frac{11}{2}$ & 5 & 1 & $ \frac{\sqrt {105}}{13} R_{54} -\frac{\sqrt {105}}{65} R_{74} -\frac{\sqrt {2730}}{455} R_{76}$\\
$\frac{5}{2}$ & 2 & 2 & $\frac{11}{2}$ & 5 & 2 & $ -\frac{5\,\sqrt {35}}{77} R_{32} +\frac{10\,\sqrt {385}}{1001} R_{52} -\frac{\sqrt {1155}}{1001} R_{54} -\frac{5\,\sqrt {210}}{2002} R_{72}$\\
 &&&&&& $\qquad  +\frac{2\,\sqrt {1155}}{715} R_{74} +\frac{3\,\sqrt {30030}}{1430} R_{76}$\\
$\frac{5}{2}$ & 2 & 2 & $\frac{11}{2}$ & 5 & 3 & $ -\frac{5\,\sqrt {70}}{231} R_{30} +\frac{10\,\sqrt {21}}{231} R_{32} +\frac{10\,\sqrt {110}}{429} R_{50} +\frac{2\,\sqrt {231}}{273} R_{52}$\\
 &&&&&& $\qquad  -\frac{\sqrt {77}}{13} R_{54} -\frac{5\,\sqrt {6}}{143} R_{70} +\frac{27\,\sqrt {14}}{1001} R_{72} -\frac{3\,\sqrt {77}}{143} R_{74}$\\
$\frac{5}{2}$ & 2 & 2 & $\frac{11}{2}$ & 5 & 4 & $ \frac{5\,\sqrt {7}}{11} R_{32} +\frac{8\,\sqrt {77}}{143} R_{52} -\frac{9\,\sqrt {231}}{1001} R_{54} -\frac{17\,\sqrt {42}}{286} R_{72}$\\
 &&&&&& $\qquad  -\frac{6\,\sqrt {231}}{1001} R_{74} -\frac{5\,\sqrt {6006}}{2002} R_{76}$\\
$\frac{5}{2}$ & 2 & 2 & $\frac{11}{2}$ & 5 & 5 & $ \frac{5\,\sqrt {35}}{33} R_{30} +\frac{5\,\sqrt {42}}{231} R_{32} -\frac{7\,\sqrt {55}}{429} R_{50} -\frac{\sqrt {462}}{3003} R_{52}$\\
 &&&&&& $\qquad  +\frac{10\,\sqrt {154}}{1001} R_{54} -\frac{42\,\sqrt {3}}{143} R_{70} -\frac{6\,\sqrt {7}}{1001} R_{72} -\frac{15\,\sqrt {154}}{1001} R_{74}$\\
$\frac{5}{2}$ & 2 & 2 & $\frac{11}{2}$ & 5 & 6 & $ \frac{50}{231}i R_{30} +\frac{5\,\sqrt {30}}{77}i R_{32} +\frac{5\,\sqrt {77}}{429}i R_{50} -\frac{3\,\sqrt {330}}{143}i R_{52}$\\
 &&&&&& $\qquad  +\frac{4\,\sqrt {105}}{715}i R_{70} -\frac{192\,\sqrt {5}}{715}i R_{72}$\\
$\frac{5}{2}$ & 2 & 2 & $\frac{11}{2}$ & 6 & 1 & $ \frac{40\,\sqrt {1023}}{4433} R_{40} -\frac{16\,\sqrt {10230}}{4433} R_{42} +\frac{28\,\sqrt {13299}}{4433} R_{60}$\\
 &&&&&& $\qquad   +\frac{72\,\sqrt {155155}}{31031} R_{62}-\frac{8\,\sqrt {17391}}{6851} R_{80} +\frac{24\,\sqrt {608685}}{47957} R_{82} $\\
 &&&&&& $\qquad   -\frac{100\,\sqrt {22134}}{47957} R_{84}$\\
$\frac{5}{2}$ & 2 & 2 & $\frac{11}{2}$ & 6 & 2 & $ -\frac{796\,\sqrt {4495}}{899899} R_{40} -\frac{1446\,\sqrt {1798}}{899899} R_{42} -\frac{216\,\sqrt {58435}}{128557} R_{60}$\\
 &&&&&& $\qquad  -\frac{2272\,\sqrt {245427}}{899899} R_{62} -\frac{12\,\sqrt {76415}}{11687} R_{80} +\frac{1200\,\sqrt {106981}}{1390753} R_{82}$\\
 &&&&&& $\qquad  +\frac{6\,\sqrt {11767910}}{15283} R_{84}$\\
$\frac{5}{2}$ & 2 & 2 & $\frac{11}{2}$ & 6 & 3 & $ -\frac{310\,\sqrt {290}}{29029} R_{40} -\frac{706\,\sqrt {29}}{29029} R_{42} -\frac{20\,\sqrt {3770}}{4147} R_{60} +\frac{8\,\sqrt {15834}}{4147} R_{62}$\\
 &&&&&& $\qquad  +\frac{4\,\sqrt {4930}}{493} R_{80} +\frac{240\,\sqrt {6902}}{44863} R_{82} -\frac{60\,\sqrt {189805}}{44863} R_{84}$\\
$\frac{5}{2}$ & 2 & 2 & $\frac{11}{2}$ & 6 & 4 & $ -\frac{84\,\sqrt {345}}{3289}i I_{41} -\frac{12\,\sqrt {4186}}{23023}i I_{61} -\frac{128\,\sqrt {10465}}{23023}i I_{63} $\\
 &&&&&& $\qquad -\frac{198\,\sqrt {2346}}{5083}i I_{81} -\frac{106\,\sqrt {301070}}{35581}i I_{83}$\\
$\frac{5}{2}$ & 2 & 2 & $\frac{11}{2}$ & 6 & 5 & $ -\frac{852\,\sqrt {2737}}{391391}i I_{41} -\frac{4\,\sqrt {152490}}{3289}i I_{61} +\frac{16\,\sqrt {15249}}{5083}i I_{63} $\\
 &&&&&& $\qquad +\frac{498\,\sqrt {1610}}{5083}i I_{81} +\frac{370\,\sqrt {1518}}{5083}i I_{83}$\\
$\frac{5}{2}$ & 2 & 2 & $\frac{11}{2}$ & 6 & 6 & $ -\frac{4\,\sqrt {510}}{221}i I_{41} -\frac{6\,\sqrt {1547}}{1001}i I_{61} +\frac{18\,\sqrt {15470}}{17017}i I_{63} +\frac{152\,\sqrt {3}}{221}i I_{81}$\\
 &&&&&& $\qquad  -\frac{24\,\sqrt {385}}{1547}i I_{83}$\\
\hline
\end{tabular}
\end{center}
\end{table}

%\section*{References}

\bibliography{cited_refs}

\end{document}